\documentclass[preprint2]{aastex}
\shorttitle{Search for megamasers in type-2 AGNs}
\shortauthors{Bennert et al.}

\begin{document}

\title{A SEARCH FOR H$_2$O MEGAMASERS IN HIGH-Z TYPE-2 AGNS}

\shorttitle{A SEARCH FOR MEGAMASERS IN TYPE-2 AGNS}

\author{Nicola Bennert\altaffilmark{1}}
\affil{Institute of Geophysics and Planetary Physics, 
University of California, 
Riverside, CA 92521}
\email{bennert@physics.ucsb.edu}
\altaffiltext{1}{Current address: Department of Physics,
University of California, Santa Barbara, CA 93106}

\author{Richard Barvainis\altaffilmark{2}}
\affil{National Science Foundation, 4301 Wilson Boulevard, Arlington, VA 22230; and Department of Physics, Gettysburg College, 300 North Washington Street, Gettysburg, PA 1732}
\email{rbarvai@nsf.gov}
\altaffiltext{2}{
Any opinions, findings, conclusions, and recommendations expressed in
this material are those of the author and do not necessarily reflect the
views of the National Science Foundation.}

\author{Christian Henkel}
\affil{Max-Planck-Institut f\"ur Radioastronomie, 
Auf dem H\"ugel 69, D-53121 Bonn, Germany}
\email{chenkel@mpifr-bonn.mpg.de}

\and

\author{Robert Antonucci}
\affil{Department of Physics, University of California, Santa Barbara, CA 93106}
\email{ski@physics.ucsb.edu}

\shortauthors{Bennert et al.}

\begin{abstract}
We report a search for H$_2$O megamasers in 274 SDSS type-2 AGNs (0.3 $<$ z $<$ 0.83), half of which can
be classified as type-2 QSOs from their 
[\ion{O}{3}]\,5007 luminosity, using the Robert C. Byrd Green Bank Telescope (GBT)
and the Effelsberg 100-m radio telescope. 
Apart from the detection of the extremely luminous water 
vapor megamaser SDSS J080430.99+360718.1, already reported by Barvainis 
\& Antonucci (2005), we do not find any additional line emission. This
high rate of non-detections is compared to
the water maser luminosity function created from the 78 water maser galaxies
known to date and its extrapolation towards 
the higher luminosities of ``gigamasers''
that we would have been able to detect given the sensitivity of our survey.
The properties of the known water masers are summarized and discussed
with respect to the nature of high-z type-2 AGNs and megamasers in general. 
In the appendix, we list 173 additional objects (mainly radio galaxies, but also QSOs and galaxies) 
that were observed with the GBT, the Effelsberg 100-m radio telescope, 
or Arecibo Observatory without leading to the detection of water maser emission. 
\end{abstract}


\section{INTRODUCTION}
The 22~GHz H$_2$O maser emission line 
is of great astrophysical interest for its extreme requirements 
for density ($>$10$^{7}$~cm$^{-3}$), temperature ($>$300~K), and of
course radial velocity coherence. 
It is detected in both Galactic and extragalactic star
forming regions as well as in the central regions of active galaxies (AGNs). 
In AGNs, isotropic luminosities 
commonly reach 
values of $L_{\rm H_2 O} > 10 L_{\odot}$ and the objects are then classified as ``megamasers'' 
(see recent reviews by e.g. \citealt{gre04,mor04,hen05b,lo05}).

So far, water megamaser emission has been detected in about 
10\% of active galactic nuclei (AGNs) surveyed in the local universe \citep{bra04}.
The association of water megamasers 
with AGNs of primarily Seyfert-2 or 
Low-Ionization Nuclear Emission-line Region
(LINER) type \citep{bra97,bra04} 
and the fact that the emission often arises from the innermost parsec(s) 
of their parent galaxy have raised great interest in the 
study of 22~GHz maser emission. It suggests that the so far poorly constrained 
excitation mechanism is closely related to AGN activity, probably 
irradiation by X-rays (e.g. \citealt{neu94}).
For Seyfert-2 galaxies, in the framework of the so-called unified model \citep{ant93},
a dusty molecular disk or torus is seen edge-on
where the conditions and velocity-coherent path lengths are favorable for the formation
of megamaser activity.

In those cases in which the emission arises from a nuclear disk 
and can be resolved spatially using
Very Long Baseline Interferometry (VLBI), the central black hole (BH) mass can be constrained,
as has been successfully shown for NGC\,4258 (e.g. \citealt{gre95,miy95,her99,her05}).
Moreover, using H$_2$O masers, distances to galaxies 
can be obtained without the use of standard 
candles  \citep{miy95,her99,arg04,bru05,arg07,hum08}.
Thus, finding new megamaser galaxies is of great interest.

If the unified scheme for AGNs is to be equally successful for objects of high as 
well as of low luminosity, 
there should exist a large number of type-2 QSOs whose optical spectra 
are dominated by narrow emission lines. 
Indeed, with the advent of new 
extended surveys such as the Sloan Digital Sky Survey (SDSS), 
many type-2 QSOs have recently been identified \citep{zak03}.

We conducted a search for water megamasers in 274 of the 
291 SDSS type-2 AGNs \citep{zak03} using the 
Robert C. Byrd Green Bank Telescope (GBT) and the
Effelsberg 100-m radio telescope.
With a redshift range of 0.3 $<$ z $<$ 0.83,
the sample covers significantly higher distances than
most previous searches for H$_2$O megamasers ($z \ll 0.1$; \citealt{bra96, 
tar03, bra04, kon06a, kon06b, bra08, cas08})
and is the first survey for water megamasers
in objects with QSO luminosities 
(except for the study of \cite{bar05} which is part of the larger
survey presented here). 

Such a search provides clues to whether the unified model can indeed be extended 
to QSOs or whether their powerful engines lead to a different scenario. 
Do the high QSO luminosities result in H$_2$O ``gigamasers''?
Or do they destroy the warm dense molecular gas needed to supply the water molecules? 
Are the molecular parts of the accretion 
disks much farther away from the nuclear engine, 
so that rotation velocities are smaller in spite of a potentially more massive 
nuclear engine than in Seyfert-2 galaxies? 
Finding megamasers in type-2 QSOs may provide insights into QSO molecular
disks and tori and enables us to independently determine their BH masses. 
Even more importantly, megamasers
in type-2 QSOs may provide the unique possibility to directly measure their distances and 
thus verify the
results from type 1a supernovae measurements on the existence and properties of 
the elusive dark energy (e.g. \citealt{bar05,bra07,rei08}).

We summarize the sample properties in \textsection{2}, describe the observations
in \textsection{3}, present the results in \textsection{4}, and discuss them
in \textsection{5}. After a brief summary (\textsection{6}),
we list a sample of 171 additional objects  
(radio galaxies, QSOs, and galaxies)
in the Appendix (\textsection{A}).
Throughout the paper, we assume a Hubble constant
of $H_0$ = 75\,km\,s$^{-1}$\,Mpc$^{-1}$.
For the high-z objects, we additionally assume
$\Omega_{\Lambda}$ = 0.73 and $\Omega_{\rm M}$ = 0.27
\citep{wri06}.

\section{SAMPLE PROPERTIES}
As already mentioned above,
our sample consists of 274 type-2 AGNs (0.3 $<$ z $<$ 0.83) selected
from the SDSS \citep{zak03}.
Out of these, 122 objects have $L_{\rm [OIII]}$ $>$ 3 10$^8$ $L_{\odot}$,
and can thus be classified as type-2 QSOs \citep{zak03}.
About 10\% of the SDSS type-2 AGNs are radio-loud \citep{zak04},
comparable to the AGN population as a whole.
A few type-2 AGNs have soft X-ray counterparts \citep{zak04}.
Spectropolarimetry was carried out for
12 type-2 QSOs and revealed polarization in all objects.
Five objects show polarized broad lines expected
in the framework of the unified model at the sensitivity achieved \citep{zak05}.
\citet{zak06} studied 
the host galaxy properties for nine objects, 
finding that the majority (6/9) of the type-2 QSO 
host galaxies are ellipticals.
All observations support the interpretation of the type-2 AGNs
selected from the SDSS as being powerful obscured AGNs.
Table~\ref{table1} summarizes the sample properties.

\section{OBSERVATIONS}
All sources were measured in the
6$_{\rm 16}$-5$_{\rm 23}$ line of H$_{\rm 2}$O (22.23508 GHz
rest frequency).
The observations were carried out during several runs at the GBT 
of the National Radio Astronomy
Observatory (NRAO) in January and June 2005 as well as at the Effelsberg 
100-m radio telescope of the 
Max Planck Institut f\"ur Radioastronomie (MPIfR) in November and December 2005. 
For details of observations, see Table~\ref{table1}.\footnote{Note that our
velocity convention is the optical one, i.e. $v$ = c$z$.}

\subsection{GBT}
A total of 128 SDSS type-2 AGNs were observed with 
the GBT, limited to those that are within
the available frequency coverage of 12--15.4 GHz (0.44 $<$ z $<$ 0.85).
The observing mode utilized two feeds separated by 5.5\arcmin\ on the sky, 
each with dual polarization. The system
temperatures were typically 25 K. The source was placed alternately in each beam, with a 
position-switching interval of 2 minutes and was typically observed for 30 minutes total
on-source time (possibly longer for objects with follow-ups).
A total of 200 MHz bandwidth was covered with
$\sim$0.5\,km\,s$^{-1}$ channels.
Antenna pointing checks were made roughly every 2 hours, and typical pointing 
errors were less than 1/10 of a full width to half power 
(FWHP) beamwidth of 48\arcsec\ at 15 GHz.
GBT flux calibration was done using standard antenna gain vs frequency curves.
We estimate the calibration uncertainty to be $\sim$20\%.

\subsection{Effelsberg}
A total of 150 SDSS type-2 AGNs were observed with 
the Effelsberg 100-m radio telescope\footnote{Note that a few objects
were observed at both GBT and Effelsberg yielding a total number
of 274 sources.}.
The measurements were carried out with a dual polarization HEMT receiver 
providing system temperatures of $\sim$36--45 K
(for the observed frequencies between $\sim$14.3 and 17.1 GHz)
on a main beam brightness temperature scale. 
The observations were obtained in a position switching mode.
Signals from individual on- and off-source positions were integrated
for 3 minutes each, with the off-position offsets alternating between
+900 and --900 arcsec in Right Ascension. 
The typical on-source integration time was $\sim$70 minutes 
(possibly longer for objects with follow-ups) with variations due to weather and elevation.
An auto-correlator backend 
was used, split into eight bands of 40\,MHz width and 512 channels, 
respectively, 
that were individually shifted in such a way that a total of 
$\sim$130--240\,MHz was covered. 
Channel spacings are $\sim$1.5\,km\,s$^{-1}$. 
The FWHP 
beamwidth was $\sim$40\arcsec.
The pointing accuracy was better than 
10\arcsec. Calibration was obtained by repeated measurements at
different frequencies toward 3C\,286, 3C\,48, and 
NGC\,7027, with flux densities taken from \citet{ott94}, interpolated
for the different observed frequencies using their Table~5.
The calibration should be accurate to $\sim$20\%. 

\section{RESULTS}
All spectra were examined carefully by eye for both broad and narrow lines.
In addition, we applied spectral binning using
several bin sizes, especially if there was anything
looking remotely like a signal.

\subsection{The Gigamaser J0804+3607}
As already reported in \citet{bar05},
water maser emission was detected from the
type-2 QSO SDSS J080430.99+360718.1 
(hereafter J0804+3607; $z$ = 0.66).
With $L_{\rm H_2O}$ $\simeq$
21,000 $L_{\odot}$\footnote{Using $H_0$ = 75\,km\,s$^{-1}$\,Mpc$^{-1}$, $\Omega_{\Lambda}$ = 0.73,
and $\Omega_{\rm M}$ = 0.27. Note that the value given by
\citet{bar05}, $L_{\rm H_2O}$ = 23,000
$L_{\odot}$, is higher due to a smaller value of $H_0$.},
it is the intrinsically most powerful maser known.

\subsection{Non Detections}
\label{non}
While the detection of a water vapor maser in J0804+3607 
shows that H$_2$O masers are indeed detectable at high redshift,
and thus, that such a project is in principle feasible,
no obvious maser emission was discovered in any of the remaining 
objects (Table~\ref{table1}).
For some objects we see 2--3 sigma blips,
which, however, do not meet the 5 $\sigma$ detection criterion.
While they are most likely statistically insignificant,
considering the effectively large number of trials
implicit in the number of channels per spectrum, and the number of
objects observed, follow-up observations are planned for verification.

\section{DISCUSSION}
Ideally, we would like to estimate the detection probabilities
for our sample and compare them with the (non-) detection rate.

In the local universe, water megamaser emission has been
detected in about 10\% of AGNs \citep{bra04}.
Simply extrapolating the percentage of megamasers detected in 
nearby Seyfert-2 galaxies 
to the more distant type-2 Seyferts and type-2 QSOs,
leads  to the expectation of finding at least $\sim$27 megamasers
among the 274 SDSS type-2 AGNs. 
However, such a naive extrapolation 
does not take into account the megamaser luminosity
function, its evolution with redshift, the sensitivity of the survey,
and intrinsic differences among the sources. We will discuss each
of these issues in turn.

\subsection{H$_2$O Maser Luminosity Function}
\label{luminosity}
\citet{hen05a} performed a statistical analysis
of 53 H$_2$O maser galaxies beyond
the Magellanic clouds.
From the maser luminosity
function (LF), i.e. the number density of objects
with a given water maser luminosity per logarithmic
interval in $L_{\rm H_2O}$, they estimate that the 
number of detectable maser sources is almost 
independent of their intrinsic luminosity:
The larger volume in which high luminosity masers can 
be detected compensates the smaller 
source density. This implies that masers out to cosmological distances 
should be detectable with current telescopes, as long as
the LF is not steepening at the very high end and if suitable candidates
are available. Thus, \citet{hen05a} conclude that most 
of the detectable luminous 
H$_2$O megamasers with $L_{\rm H_2 O} > 100 L_{\odot}$ 
have not yet been found.

\begin{figure}
\includegraphics[scale=0.29,angle=-90]{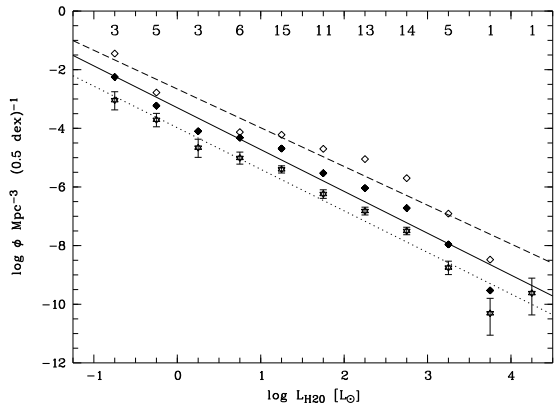}
\caption{Luminosity function for water maser galaxies 
at $D$ $\ge 100$\,kpc
 (cf. Table~\ref{maserlf}). Plotted are the resulting LFs 
assuming three different sensitivity limits of the survey:
1 Jy km\,s$^{-1}$ (open diamonds), 0.2 Jy km\,s$^{-1}$ (filled diamonds),
and 0.06 Jy km\,s$^{-1}$ (open stars). 
The numbers on the top indicate the number of galaxies included in each
luminosity bin for a sensitivity limit of 0.06 Jy km\,s$^{-1}$
(open stars);
corresponding error bars were calculated from Poisson statistics \citep{con89}.
The lines indicate the best linear fit for a sensitivity
limit of 1 Jy km\,s$^{-1}$ (dashed line; $\Phi \propto L_{\rm H_2O}^{-1.3}$), 0.2 Jy km\,s$^{-1}$ 
(solid line; $\Phi \propto L_{\rm H_2O}^{-1.4}$),
and 0.06 Jy km\,s$^{-1}$, respectively (dotted line; $\Phi \propto L_{\rm H_2O}^{-1.4}$).
See text for further details.}
\label{lf}
\end{figure}

We performed a similar analysis of the larger sample
of masers known to date (78 sources; see Table~\ref{maserlf})
and derived a zeroth order approximation of the LF of extragalactic
water maser sources.
We here briefly summarize the procedure adapted from \citet{hen05a};
for details and a discussion of limitations, we refer the reader to
\citet{hen05a}.
To estimate the water maser LF, the standard $V/V_{\rm max}$ method
\citep{sch86} was used. We divided the 78 maser sources known to date
(Table~\ref{maserlf}) in luminosity bins $L_b$ of 0.5 dex 
($b$ = 1,...,11), covering a total range of
$L_{\rm H_2 O}/L_{\odot} = 10^{-1}$ to $3\,10^{4}$.
The differential LF value was calculated for each luminosity bin according to
\begin{eqnarray*}
\Phi (L_b) = \frac{4 \pi}{\Omega} \Sigma^{n(L_b)}_{i=1} (1/V_{\rm max})_i\hspace*{0.15cm} .
\end{eqnarray*}
$n(L_b)$ is the number of galaxies with $L_b - 0.25 < \log (L_{\rm H_2O}/L_{\odot}) \le L_b + 0.25$ (centering 
on $\log (L_{\rm H_2O}/L_{\odot})$ = -0.75, -0.25, +0.25, etc.).
Following \citet{hen05a}, 
we set $\Omega$ = 2 $\pi$, approximating the sky coverage to be the entire northern sky,
for the Seyfert sample. For J0804+3607, we assumed
 $\Omega$ = 0.64 as the SDSS data release 1 from which the type-2 AGN sample of \citet{zak03}
was taken covered $\sim$2100\,deg$^2$.
$V_{\rm max}$ is the maximum volume over which an individual galaxy can be 
detected depending on the detection limit of the survey and its maser luminosity
(see also Sect.~\ref{sensitivity}).
We calculated the maser LF for three different detection limits: 
(a) 1 Jy km\,s$^{-1}$, (b) 0.2 Jy km\,s$^{-1}$, and (c) 0.06 Jy km\,s$^{-1}$.
The first two cases are identical to the procedure in \citet{hen05a},
the latter case was added to include objects such as the gigamaser J0804+3607\footnote{Note that for this distant object, we used the co-moving volume as maximum volume.}.
From the sample of 78 sources, IC\,342 is excluded in all three cases
due to its too low maser luminosity. 
In case (a), 32 masers fall below the chosen detection limit,
and in case (b), 10 galaxies were omitted.
In case (c), all 77 sources are included in the LF.
The resulting LFs are shown in Fig.~\ref{lf}.

The overall slope of the H$_2$O LF does not depend strongly
on the chosen detection limit.
Applying a linear fit to the three different LFs, 
we derive $\Phi \propto L_{\rm H_2O}^{-1.4 \pm 0.1}$,
comparable to \citet{hen05a}, but steeper than the LF
for OH megamasers ($\Phi \propto L_{\rm H_2O}^{-1.2}$) \citep{dar02}.
The main conclusions we can draw from this new version of the water maser
LF are virtually identical to those by \citet{hen05a}:
(i) The number of sources at the upper end of the LF decays rapidly,
indicating that gigamasers are intrinsically
rare or that the proper sources have not
yet been found --- so far most surveys were focused on nearby sources.
In case c, when including J0804+3607, the LF seems to raise again
which is due to the much smaller area of sky covered in the survey
presented here (see above).
(ii) There are only a few sources in the 
$L_{\rm H_2O}$ = 0.1 -- 10 $L_{\odot}$ bins.
The associated slight minimum in the LF suggests that 
two different LFs are overlayed:
one for masers in star forming regions with low
luminosities ($L_{\rm H_2O}$ $<$ 0.1$-$10 $L_{\odot}$)
and one for maser sources in AGNs with $L_{\rm H_2O}$ $>$
10 $L_{\odot}$.

However, note that an extrapolation of the local
maser LF to higher redshifts is not straightforward.
It would assume no cosmological evolution, but a strong
evolution of AGN activity with redshift is known.
Another cautionary note we want to add is that 
our survey is most sensitive to narrow-line masers
and that we might be missing broad-line masers.
Although we binned our data in various ways to emphasize
potential broad-line masers and to make them more visible, a given amount
of integrated flux density would then be spread over a larger amount of
noise and baseline uncertainties would become more severe.
Broad lines typically
arise in jet masers such as Mrk\,348 \citep{pec03} and NGC\,1052
\citep{cla98},
one exception being TXS\,2226-184 where a broad maser arises
from a disk maser \citep{bal05}; for a discussion on jet masers
see also \citet{hen05b}.
As these broad-line masers are included in the LF of the
known maser sources, we in principle introduce a systematic
error when extrapolating the derived LF
to our survey. However, since broad-line masers seem to be rare,
we neglect this problem.

\subsection{Sensitivity of the Survey}
\label{sensitivity}
We can estimate the H$_2$O luminosities we would 
be able to detect
depending on the sensitivity of our survey.
Our sample lies at a redshift range of 0.3 $<$ $z$ $<$ 0.83,
corresponding to luminosity distances of $D_L$ = 1,460 $-$ 4,980 
Mpc\footnote{Using $H_0$ = 75\,km\,s$^{-1}$\,Mpc$^{-1}$, 
$\Omega_{\lambda}$ = 0.73,
and $\Omega_{\rm matter}$ = 0.27}.
The detectable H$_2$O luminosities depend on the sensitivity of the 
survey and the distance of the object \citep{hen05a}.
In general, it is 
\begin{eqnarray*}
L = \frac{F_{\nu^{\prime}} (\nu_0)}{1+z} \times 4 \pi D_L^2\hspace*{0.15cm} ,
\end{eqnarray*}
with the specific flux $F_{\nu^{\prime}} (\nu_0)$ in the observed frame.
Then
\begin{eqnarray*}
\frac{L_{\rm H_2O}}{L_{\odot}}  =
\frac{10^{-23}\,S_{\rm peak}}{\rm Jy}
\times \frac{\nu_{\rm rest}}{\rm{c}}
\times \frac{\Delta v}{{\rm km\,s^{-1}}}
\times \frac{1}{1+z}\\
\times 4 \pi  \left(\frac{3.1\,10^{24}\,D_L}{{\rm Mpc}}\right)^2 
\times \frac{1}{3.8\,10^{33}}\hspace*{0.15cm} ,
\end{eqnarray*}
where $\nu_{\rm rest}$ = 22.23508 GHz and $c$ the speed of light in km\,s$^{-1}$.
Thus
\begin{eqnarray*}
\frac{L_{\rm H_2O}}{L_{\odot}}  = 
\left[0.023 \times \frac{S_{\rm peak}}{{\rm Jy}} 
\times \frac{\Delta v}{{\rm km\,s^{-1}}}\right]
\times
\frac{1}{1+z} \times \left(\frac{D_L}{{\rm Mpc}}\right)^2 
\end{eqnarray*}
(see also \citealt{sol05}).
Assuming a characteristic linewidth of the dominant spectral feature
of 20\,km\,s$^{-1}$, a 5 $\sigma$ detection threshold of 
5$\times$(7.6/4.5)\,mJy 
(with 7.6 mJy being the average rms of our observations for a 1\,km\,s$^{-1}$ channel)
gives
\begin{eqnarray*}
\frac{L_{\rm H_2O}}{L_{\odot}}  
&=& 
0.0039
\times \frac{1}{1+z} \times
\left(\frac{D_L}{{\rm Mpc}}\right)^2 \hspace*{0.15cm} .
\end{eqnarray*}
Thus, given the distance of our sample,
we can detect H$_2$O luminosities of $L_{\rm H_2O}$ 
$\simeq$ 6,400 $-$ 52,900  $L_{\odot}$.

These H$_2$O luminosities are  higher
than the average luminosity found for megamasers in
Seyfert-2 galaxies and LINERs. 
Among the 78 known H$_2$O maser galaxies,
the typical cumulative H$_2$O
luminosity range is 10 ... 2000 L$_{\odot}$ for sources
associated with AGNs, while
most of the weaker masers appear to be related to
star formation.
However, in addition, there are two gigamasers known,
TXS\,2226--184 \citep{koe95} with  $L_{\rm H_2O}$ = 6800 $L_{\odot}$
and J0804+3607 with $L_{\rm H_2O}$ $\simeq$
21,000 $L_{\odot}$.
Thus, the distance of our sample allows us to detect gigamasers
comparable to TXS\,2226--184 and J0804+3607 only.
The low detection rate may simply reflect that
megamasers with H$_2$O luminosities 
above 6,000 $L_{\odot}$ are intrinsically rare,
an interpretation that is supported by the water
maser LF (Sect.~\ref{luminosity}).
However, there are other possibilities for the low
detection rates that we discuss in the following.

\subsection{Velocity Coverage}
Our observations cover a frequency width of 130--240 MHz, 
corresponding to $\sim$1800--4000 km\,s$^{-1}$.
This range should be large enough to cover
any mismatch in redshift between the maser emission
and the optical [OII]\,$\lambda$3727 emission.
For J0804+3607, for example, the megamaser line
is redshifted with respect to the [OII] line by
360\,km\,s$^{-1}$ \citep{bar05}.
However, we may not be able to detect 
superpositions of thousands of individual maser components
with slightly differing velocities
nor rapidly rotating tori with only the tangential parts showing
strong (highly red- and blue-shifted) maser emission \citep{hen98},
if the emission covers a range of $>$2000\,km\,s$^{-1}$.

\subsection{Time Variability}
Monitoring of megamaser sources has revealed
variability on timescales of weeks with fluctuations
of the order of 10\% (e.g. \citealt{gre97b}) as well
as on timescales of years with maser luminosities
varying by factors of 3-10 (e.g. \citealt{fal00a,gal01,tar07}).
Such flaring masers can
be explained by an increase in the X-ray luminosity
of the AGN \citep{neu00}, if the maser emission
is powered by the X-ray radiation from the AGN.

We cannot exclude that at least some of the sources
for which we did not detect megamaser emission
were in a low stage of maser activity
and might be detected at a later flaring stage.

\subsection{Intrinsic Differences}
So far, we did not take into account that,
when comparing the low-luminous AGNs such as Seyfert-2
galaxies and LINERs with the high-luminous AGNs such
as the type-2 QSOs in our sample,
we may be comparing apples and oranges.
Intrinsic differences between the different samples
complicates estimating detection probabilities.

\subsubsection{The Nature of Megamaser Galaxies}
So far, $\sim$1500 galaxies have been searched for H$_2$O maser emission,
resulting in the detection of 78 maser 
galaxies (Table~\ref{maserlf}).
For 73 of the 78 known H$_2$O maser galaxies,
the activity type has been determined (NED\footnote{Note that NED classifications
such as morphological types and activity classes are inhomogeneous.}; see Table~\ref{maserlf}).
The vast majority are classified as Seyferts (78\%), out of which
Sy2s (including Sy1.9) are the dominant type (88\%) and Sy1s are rare (3\%),
the rest being classified simply as Sy, or Sy1.5. 
The second largest activity type among the extragalactic water
maser sources are LINERs, making up 11\% of the sample.
In addition, 7\% are HII regions, 3\% starburst (SB) galaxies and 1\% Narrow-Line Radio Galaxies
(NLRGs)\footnote{Note that we counted the ``more energetic'' activity type,
e.g. an object with activity types ``Sy2, SB, HII'' (Table~\ref{maserlf},
column 10), was counted as Sy2, an object with ``L, LIRG, HII'', was counted
as LINER, etc.}. 

Objects with activity type of HII or SB have generally 
lower maser luminosities and fall in the kilomaser range ($L_{\rm H_2O}$ $<$
10 $L_{\odot}$).
When including only maser sources with $L_{\rm H_2O} \ge 10 L_{\odot}$,
the activity type has been determined for 57 sources in total.
Out of these, 86\% are Seyferts (82\% are Sy2s; 4\% are Sy1s, namely NGC\,235A
and NGC\,2782), 10\% are LINERs
and only 2\% HII regions 
(namely NGC\,2989).

Using the higher number of extragalactic
water maser sources known to date, our statistic thus
confirms earlier studies that
water megamaser sources are associated 
with AGNs of primarily Seyfert-2 or 
LINER type \citep{bra97,bra04}. This in turn
strengthens the general expectation to find megamasers
also in type-2 QSOs.

Interpreting this finding in the framework of the unified
models of AGNs,
where an optically thick obscuring
dust torus is envisioned to encircle the accretion disk
and type-1 AGNs are seen pole-on while type-2 AGNs are seen edge-on \citep{ant93},
suggests that the megamaser activity is related to the 
large column densities of molecular gas along the line-of-sight
in the torus.
However, even if such an interpretation holds,
the question remains why not all type-2 AGNs are megamasers.
What are the necessary ingredients for the occurrence of these 
powerful masers? 

\citet{bra97} addressed this question by a statistical
comparison of the physical, morphological,
and spectroscopic properties of the known megamaser
galaxies with those of non-megamaser galaxies\footnote{Here and in the following, we
denote as ``non-megamaser galaxies''
those galaxies that have been observed at 22~GHz, but for which no
megamaser emission was detected.}.
They compared the AGN class, the host galaxy type and inclination,
the mid-infrared (MIR) and FIR properties, 
the radio fluxes and luminosities, the [OIII]
fluxes and luminosities, and the X-ray properties
of the 16 megamasers known at that time 
with those of $\sim$340 non-megamaser galaxies. 
Apart from their main conclusion that H$_2$O emission
is only detected in Seyfert-2 galaxies and LINERs
but not in Seyfert-1 galaxies
(a conclusion that still holds for the larger
sample of megamasers known today; see above), \citet{bra97}
found that H$_2$O emission is preferentially
detected in sources that, when compared to the
non-megamaser galaxies in their sample, are 
``apparently brighter at MIR and FIR and centimeter radio wavelengths''
However, this result may at least in part result from the fact that
the megamaser galaxies are nearer than the non-megamaser galaxies.
\citet{bra97} also find that H$_2$O emission is preferentially
detected in sources with high X-ray-absorbing columns of gas --
a result that is still discussed controversially:
While \citet{zha06} 
concluded that  H$_2$O 
megamasers have similar X-ray absorbing column densities as other Seyfert-2 galaxies,
\citet{gre08} find a correlation between maser emission
and high X-ray obscuring columns.

The requirement for velocity coherence may play an
important role for the (non) occurrence of megamasers.
For NGC\,4258, for example, the scattered light requires
a thick obscuring disk (in terms of its optical shadowing
properties) but the masers reside in a thin disk \citep{wilk95,bar99,hum08}. 
Enough velocity coherence (and gas column density) is perhaps
achieved most often in the midplane. Thus, the solid angle into which
the water maser emission is beamed is small, much smaller
than that of the torus. With such a small angle,
the likelihood to observe
a maser line depending on the viewing angle is small as well.

However, now, over 10 years after the study of \citet{bra97},
the number of known megamaser galaxies has 
more than quadrupled. But there is no comparable study 
addressing the IR properties of megamasers, their host galaxies
and their radio and optical properties.
Such a study might reveal the necessary ingredients for
the occurrence of megamasers in AGNs. This in turn 
would greatly facilitate the pre-selection of
promising candidates for megamaser
emission among type-2 QSOs.
However, such a detailed comparison is beyond the scope of this paper. 

\subsubsection{FIR Luminosities and Dust Temperatures}
\begin{figure}[h!]
\includegraphics[scale=0.29,angle=-90]{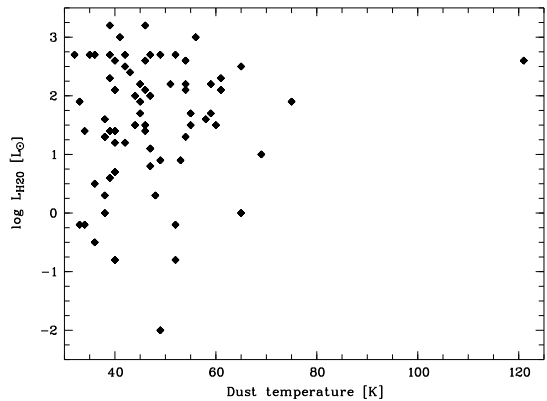}
\caption{Water maser luminosities versus dust temperatures
of H$_2$O detected galaxies (cf. Table~\ref{maserlf};
excluding those objects for which we do not
have dust temperatures, leaving us with 73 objects total). }
\label{dust}
\end{figure}

\begin{figure}[h!]
\includegraphics[scale=0.29,angle=-90]{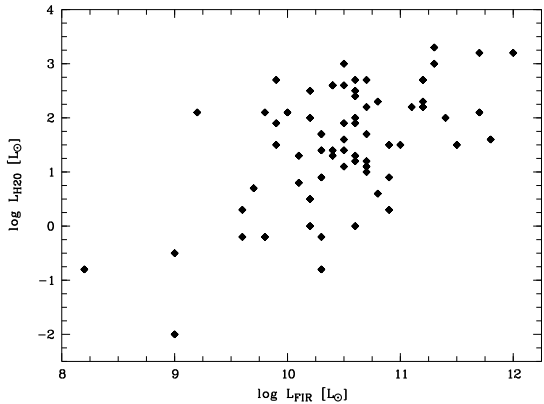}
\caption{IRAS point source FIR luminosity 
versus total H$_2$O luminosity of H$_2$O detected galaxies
(cf. Table~\ref{maserlf}; excluding those objects for which we do not
have FIR luminosities as well as those for which we only
have lower limits, leaving us with 71 objects total).}
\label{fir}
\end{figure}

Here, we derived the FIR luminosities and dust temperatures
from IRAS fluxes \citep{ful89} using the procedure of 
\citet{wou86}.
Some caution is required because these IRAS measurements are affected by the ratio
of the contribution of nuclear light to host light which depends on nuclear
FIR luminosity, nature of the host, and metric aperture size (and thus distance).
Table~\ref{maserlf} gives FIR luminosities and dust temperatures
for the sample of known maser sources.
The latter are all well above 30 K and thus rather large, as already
noted by \citet{hen86,bra97,hen05a}.\footnote{Note that the dust temperatures
were calculated from the 60/100$\mu$m flux ratio. Cooler dust might be dominant in these galaxies
but does not radiate at these wavelengths.}
There is no obvious relation between dust temperatures 
and $L_{\rm H_2O}$ (Fig.~\ref{dust}).
In Fig.~\ref{fir}, we show the FIR luminosity versus water
maser luminosity.
There seems to be a correlation between FIR luminosity
and water maser luminosities in the sense that
higher FIR luminosity lead to higher water maser luminosities.
However,
we do not claim that the maser luminosity versus FIR luminosity
plot shows an intimate physical connection between both properties.
Instead, Fig.~\ref{fir} mainly shows the range of FIR and H$_2$O
luminosities covered by the known megamaser sources.

Unfortunately, for our sample of 274 SDSS type-2 AGNs, IRAS fluxes 
are only available for 8 objects (see Table~\ref{zakfir}).
Keeping in mind the small number statistics,
it is interesting to note that the average
dust temperature for these 8 objects, $T_{\rm dust, ave}$ $\simeq$ 36$\pm$1\,K,
is by $\sim$12\,K lower 
than the one for the known maser galaxies 
($T_{\rm dust, ave}$ $\simeq$ 48$\pm$2\,K).
At the same time, the type-2 AGNs of the SDSS sample all have
very high FIR luminosities [$\log (L_{\rm FIR}/L_{\odot}) \simeq 13.5 \pm 0.1$]
compared to the maser sources [$\log (L_{\rm FIR}/L_{\odot}) \simeq 10.5 \pm 0.1$],
which are of course a lot closer.

\subsubsection{BH Mass and Accretion Disk}
One difference between the high-z type-2 AGNs in our
sample and the local Seyfert-2 galaxies and LINERs
in which megamasers have been found is the mass
of the central engine. 
For Seyfert galaxies, BH masses range between
$\sim$10$^6$ $M_{\odot}$ and a few 10$^7$ $M_{\odot}$ 
(e.g. \citealt{gre97b, her99, hen02})
while masses in QSOs can reach 10$^9$ $M_{\odot}$ or more
(e.g. \citealt{lab06, ves08}).
\citet{tar07} suggest that clouds in a disk with large rotational
velocity and small galactocentric radius like NGC\,4258 might not
be stable in the vicinity of such a large BH mass.

\subsubsection{Dust Torus and X-ray Luminosity}
\citet{bar05} have argued that extremely powerful
masers might be expected from high-luminosity QSOs.
With every square parsec of area illuminated by the
primary AGN x-ray emission, a luminosity of $\sim$100 $L_{\odot}$
is produced \citep{neu94}.
The area of the torus illuminated by the AGN increases with 
the optical/UV continuum luminosity as the
dust sublimation radius scales as $L_{\rm opt/UV}^{1/2}$.
Indeed, a scenario in which the high QSO
luminosities result in gigamasers is consistent with
the detection of J0804+3607, by far the
most powerful maser known today.
If $r_{\rm sub}$ increases with optical/UV luminosity,
also the molecular parts giving rise to the maser 
emission are expected to arise further away
from the nuclear engine. Such a prediction can be tested
observationally: We would expect to observe rotation velocities that 
are smaller than in Seyfert-2 galaxies, despite a potentially more massive BH.

However, these considerations do not take into account
that the dust covering factor may be decreasing with
optical luminosity \citep{sim05}.
In addition, the water maser luminosity is expected
to grow more slowly than optical luminosity,
since $L_{\rm x-ray}$/$L_{\rm opt}$ seems to be a declining function 
with increasing luminosity \citep{vig03}. These two effects may
cause megamasers to be intrinsically rare among type-2 QSOs.

\subsubsection{Host Galaxy}
One known difference between QSOs and Seyfert galaxies are their host galaxies:
While the majority of Seyfert galaxies reside in spiral-like galaxies,
QSOs are found predominantly in early-type galaxies
\citep[e.g.,][]{dis95,bah97,mcl99,flo04}.

Among the 78 known H$_2$O maser galaxies, the host galaxy properties
have been determined for 74 objects (NED\footnote{As noted above, the 
morphological types given by NED are inhomogeneous.}; see Table~\ref{maserlf}). 
The majority
of known megamasers resides in spiral galaxies ($\sim$ 84\%),
of which more than half were classified as barred or at least weakly barred galaxies
(53\%).
7\% of the host galaxies were classified as SO, and 
only 1\% as elliptical galaxies, the rest as irregular or peculiar galaxies ($\sim$ 8\%).
It is remarkable that 
only one galaxy, namely NGC\,1052, has an elliptical morphology.
Also, the search for megamaser emission
from early-type galaxies with FRI radio morphology
lead to no detection \citep{hen98}.

Does this imply that spiral galaxies somehow
favor the presence of megamaser activity?
Elliptical galaxies may simply lack the molecular gas necessary
for the occurrence of H$_2$O masers.
With respect to the nuclear activity,
one important difference between spiral galaxies and early-type
galaxies seems to be the fueling mechanism.
While there is now convincing evidence that most if not all QSOs
are triggered by mergers (e.g., \citealt{hut94,can01,guy06,can07,ben08,urr08}),
their low-luminous cousins, Seyfert galaxies, do not show
unusually high rates of interaction \citep[e.g.][]{mal98}.
For Seyferts, the gas necessary for the fueling of the AGN
may simply be provided by their spiral host galaxies and funneled
into the very center through
bar instabilities \citep[]{com06}.\footnote{However, the axis of the 
spiral disk seems to be
completely uncorrelated with that of the accretion disk as traced
by the radio jet (e.g. \citealt[]{sch02}).}.
Bars may also
play an important role for the obscuration of the central AGN \citep{mai99}. 
Is the fueling via bar instabilities a more stable mechanisms
ensuring the existence of a central dusty region in which the water molecules
can survive? 
And are these regions destroyed by the more violent process of fueling
by mergers?
If this is the case, one might expect to find megamasers preferentially
in barred galaxies.
However, the percentage of barred galaxies in the sample of known megamasers
residing in spiral galaxies is with 54\% (see above)
not higher than what is
typically found for the fraction of barred galaxies in the local
universe (e.g \citealt{bar08}).

\section{SUMMARY}
We report a search for megamasers in 274 SDSS type-2 AGNs (0.3 $<$ $z$
$<$ 083), half of which are luminous enough (in [OIII]) to be classified
as type-2 QSOs \citep{zak03}.
Apart from the detection of the gigamaser J0804+3607 
already reported by \citet{bar05}, we do not find any additional line emission.
We estimate the detection probabilities by comparing
our sample with known megamasers, taking into account
the observed H$_2$O maser luminosity function and
the sensitivity of our survey. 
We discuss intrinsic differences between the known
megamasers, mainly low-luminous AGNs such as Seyfert-2 galaxies
and LINERs in the local universe, and our sample consisting
of high-luminous AGNs at higher redshift.
At this stage, we cannot distinguish between the different
scenarios presented that could lead to the high rate of non-detections.

Further and more sensitive observations 
are required, e.g. using the Square Kilometer Array (SKA).
Detecting megamasers in type-2 QSOs remains a challenging
and yet, if successful, highly rewarding project
not only to determine BH masses but especially
for its possibility to constrain distances and thus
the properties of the elusive dark energy.

\acknowledgments
We thank Neil Nagar for his help with
the luminosity function.
We thank Phil Perillat and Chris Salter
for help with the Arecibo observations and reductions.
We thank the anonymous referee for carefully reading
the manuscript and for useful suggestions. 
N.B. is supported through
a grant from the National Science Foundation (AST 0507450).
The National Radio Astronomy Observatory is a 
facility of the National Science Foundation operated under 
cooperative agreement by Associated Universities, Inc.
The 100-m radio telescope at
Effelsberg is operated by the Max-Planck-Institut f{\"u}r Radioastronomie
(MPIfR) on behalf of the Max-Planck-Gesellschaft (MPG).
The Arecibo Observatory is part of the National Astronomy and Ionosphere Center, 
which is operated by Cornell University under a cooperative agreement with the National Science Foundation.
This research has made use of the NASA/IPAC Extragalactic Database (NED) 
which is operated by the Jet Propulsion Laboratory, California Institute of 
Technology, under contract with the National Aeronautics 
and Space Administration.
For the reduction and analysis of the Effelsberg data,
we used the GILDAS/CLASS software (http://www.iram.fr/IRAMFR/GILDAS).

\begin{deluxetable}{lccccccccc}
\tabletypesize{\scriptsize}
\tablecolumns{10}
\tablewidth{0pc}
\tablecaption{Details of Observations: 274 SDSS Type-2 AGNs}
\tablehead{
\colhead{Source} & \colhead{$z$} & \colhead{log $L_{\rm [OIII]}$} & \colhead{FIRST}
& \colhead{$\nu$} & \colhead{rms}  & \colhead{$v$ range} & Channel & \colhead{Telescope} & \colhead{Epoch}\\
& & & & & & & width &\\
& & ($L_{\odot}$) & (mJy) & (MHz) & (mJy) & (km\,s$^{-1}$) & (km\,s$^{-1}$) & \\
\colhead{(1)} & \colhead{(2)} & \colhead{(3)}  & \colhead{(4)} & \colhead{(5)}
& \colhead{(6)} & \colhead{(7)} & \colhead{(8)} & \colhead{(9)} & \colhead{(10)}}
\startdata
{\bf 080430.99+360718.1}& 0.658 & 8.83 & 76.96 & 13410.80 & 2.0  & --2235,2235  & 0.5 & GBT 	   & 0105\\   
002531.46--104022.2 & 0.303 &  8.73  & 1.41    & 17064.53 & 9.1  & --520,1150   & 1.4 & Effelsberg & 1105\\
002711.90+002231.8  & 0.437 &  7.92  &         & 15473.30 & 1.9  & --1935,1935  & 0.5 & GBT	   & 0605\\
002827.78--004218.8 & 0.418 &  8.75  &         & 15680.59 & 6.0  & --520,1320   & 1.5 & Effelsberg & 1105\\
002852.87--001433.6 & 0.310 &  8.43  &         & 16973.34 & 18.8 & --1650,1600  & 1.4 & Effelsberg & 1205\\
004020.31--004033.5 & 0.568 &  8.25  & 97.70   & 14180.50 & 1.9  & --2110,2110  & 0.5 & GBT	   & 0605\\
004412.87+003606.8  & 0.502 &  8.27  &	       & 14803.60 & 2.1  & --2025,2025  & 0.5 & GBT	   & 0105\\
005515.82--004648.6 & 0.345 &  8.15  &	       & 16533.17 & 9.5  & --1630,1710  & 1.4 & Effelsberg & 1205\\
005621.72+003235.8  & 0.484 &  9.45  &  8.60   & 14983.20 & 2.0  & --2000,2000  & 0.5 & GBT	   & 0105\\
005733.95+001248.3  & 0.377 &  7.71  &	       & 16147.48 & 8.6  & --1680,1750  & 1.5 & Effelsberg & 1205\\
005824.02+005153.3  & 0.347 &  7.68  &	       & 16507.00 & 11.9 & --1640,1710  & 1.4 & Effelsberg & 1205\\
011228.08--010058.2 & 0.388 &  7.98  &	       & 16019.51 & 9.9  & --1680,1760  & 1.5 & Effelsberg & 1205\\
011429.61+000036.7  & 0.389 &  8.66  &	       & 15996.46 & 7.8  & --1690,1750  & 1.5 & Effelsberg & 1205\\
011522.19+001518.5  & 0.390 &  8.14  &	       & 15996.46 & 10.8 & --1700,1760  & 1.5 & Effelsberg & 1205\\
012032.21--005502.0 & 0.601 &  9.28  &	       & 13888.30 & 2.2  & --2155,2155  & 0.5 & GBT	   & 0105\\
012325.57+001032.0  & 0.341 &  7.88  &	       & 16580.97 & 10.3 & --1630,1700  & 1.4 & Effelsberg & 1205\\
012341.47+004435.9  & 0.399 &  9.13  & 11.83   & 15893.55 & 4.7  & --500,1290   & 1.5 & Effelsberg & 1105\\
013401.80+000845.0  & 0.418 &  8.86  &         & 15680.59 & 7.4  & --520,1310   & 1.5 & Effelsberg & 1105\\
013416.34+001413.6  & 0.555 &  9.53  &	       & 14299.10 & 2.0  & --2095,2095  & 0.5 & GBT	   & 0105\\
013801.57--004946.5 & 0.433 &  8.29  &	       & 15516.50 & 2.2  & --1930,1930  & 0.5 & GBT	   & 0605\\
013856.14+003437.4  & 0.478 &  8.29  &	       & 15044.00 & 2.3  & --1990,1990  & 0.5 & GBT	   & 0105\\
013947.31--004305.1 & 0.443 &  8.00  &	       & 15408.90 & 2.0  & --1945,1945  & 0.5 & GBT 	   & 0605\\
014237.49+144117.9  & 0.389 &  8.76  & n/a     & 16007.98 & 3.8  & --500,1290   & 1.5 & Effelsberg & 1105\\
014401.31--003220.3 & 0.599 &  8.91  &	       & 13905.60 & 1.9  & --2155,2155  & 0.5 & GBT 	   & 0105\\
                    &       &        &         & 13905.60 & 2.0  & --2155,2155  & 0.5 & GBT 	   & 0605\\
014612.80+005112.4  & 0.622 &  8.51  &  1.61   & 13708.40 & 2.1  & --2185,2185  & 0.5 & GBT 	   & 0605\\
014657.24+005537.2  & 0.422 &  8.04  &	       & 15636.48 & 12.4 & --1730,1800  & 1.5 & Effelsberg & 1205\\
014828.93--093840.6 & 0.425 &  7.73  &	       & 15603.56 & 12.9 & --1740,1800  & 1.5 & Effelsberg & 1205\\
014932.53--004803.7 & 0.566 &  9.36  &  1.03   & 14198.60 & 1.9  & --2110,2110  & 0.5 & GBT 	   & 0105\\
015716.92--005304.8 & 0.422 &  9.52  &         & 15636.48 & 6.0  & --520,1330   & 1.5 & Effelsberg & 1105\\
015813.92--005306.4 & 0.545 &  8.93  &	       & 14391.60 & 1.9  & --2080,2080  & 0.5 & GBT 	   & 0605\\
015911.66+143922.5  & 0.319 &  8.56  & n/a     & 16857.53 & 18.4 & --1680,1610  & 1.4 & Effelsberg & 1205\\
020234.56--093921.9 & 0.302 &  8.12  &	       & 17077.63 & 21.1 & --1640,1580  & 1.4 & Effelsberg & 1205\\
021047.01--100152.9 & 0.540 &  9.79  &	       & 14438.40 & 1.8  & --2075,2075  & 0.5 & GBT 	   & 0105\\
                    &       &        &         & 14438.40 & 2.0  & --2075,2075  & 0.5 & GBT 	   & 0605\\
021059.66--011145.5 & 0.384 &  8.10  &	       & 16065.81 & 11.3 & --1700,1750  & 1.5 & Effelsberg & 1205\\
021757.82--011324.4 & 0.375 &  8.55  &  1.60   & 16170.97 & 12.9 & --1680,1730  & 1.4 & Effelsberg & 1205\\
021758.19--001302.7 & 0.344 &  8.75  &  1.75   & 16543.96 & 13.7 & --1650,1690  & 1.4 & Effelsberg & 1205\\
021834.42--004610.3 & 0.372 &  8.85  &	       & 16206.33 & 12.9 & --1680,1720  & 1.4 & Effelsberg & 1205\\
021910.76+005919.4  & 0.346 &  8.09  &	       & 16519.38 & 13.1 & --1630,1700  & 1.4 & Effelsberg & 1205\\
021913.61+000926.5  & 0.350 &  7.87  &	       & 16470.43 & 12.6 & --1650,1710  & 1.4 & Effelsberg & 1205\\
022004.64+005908.3  & 0.413 &  8.26  &	       & 15736.08 & 13.0 & --1720,1790  & 1.5 & Effelsberg & 1205\\
022028.74--000641.2 & 0.364 &  7.44  &	       & 16301.38 & 13.2 & --1680,1730  & 1.4 & Effelsberg & 1205\\
022049.89+001845.8  & 0.398 &  8.04  &	       & 15904.92 & 11.0 & --1710,1770  & 1.5 & Effelsberg & 1205\\
022146.36--004511.9 & 0.357 &  7.95  &	       & 16385.47 & 12.1 & --1680,1710  & 1.4 & Effelsberg & 1205\\
022214.12+004527.5  & 0.421 &  8.37  &	       & 15647.49 & 13.1 & --1730,1800  & 1.5 & Effelsberg & 1205\\
022234.58--002902.2 & 0.350 &  7.61  &	       & 16470.43 & 12.9 & --1650,1710  & 1.4 & Effelsberg & 1205\\
022341.02+011446.6  & 0.307 &  7.70  & 14.77   & 17012.30 & 21.1 & --1660,1580  & 1.4 & Effelsberg & 1205\\
022344.01+003914.5  & 0.397 &  8.25  &	       & 15916.31 & 11.0 & --1700,1770  & 1.5 & Effelsberg & 1205\\
022453.74+002031.6  & 0.385 &  7.82  &	       & 16054.21 & 11.8 & --1700,1760  & 1.5 & Effelsberg & 1205\\
022459.01--004719.6 & 0.327 &  7.46  &	       & 16755.90 & 26.2 & --1680,1600  & 1.4 & Effelsberg & 1205\\
022606.86--001656.0 & 0.407 &  8.16  &	       & 15803.18 & 13.0 & --1730,1780  & 1.5 & Effelsberg & 1205\\
022652.76--001200.8 & 0.408 &  8.08  &	       & 15791.96 & 11.9 & --1730,1780  & 1.5 & Effelsberg & 1205\\
022701.23+010712.3  & 0.363 &  8.90  &	       & 16313.34 & 11.8 & --1660,1710  & 1.4 & Effelsberg & 1205\\
022728.49+005045.1  & 0.306 &  8.18  & 12.47   & 17025.33 & 22.6 & --1640,1590  & 1.4 & Effelsberg & 1205\\
023359.93+004012.7  & 0.388 &  8.17  &	       & 16019.51 & 10.6 & --1700,1740  & 1.5 & Effelsberg & 1205\\
023411.77--074538.4 & 0.310 &  8.77  &  2.47   & 16973.34 & 7.5  & --510,1140   & 1.4 & Effelsberg & 1105\\
023759.76+001723.6  & 0.335 &  7.46  &	       & 16655.49 & 9.6  & --1690,1610  & 1.4 & Effelsberg & 1205\\
024240.92+004612.1  & 0.408 &  8.25  &  1.98   & 15791.96 & 11.9 & --1720,1780  & 1.5 & Effelsberg & 1205\\
024309.79+000640.3  & 0.414 &  7.95  &	       & 15724.95 & 13.0 & --1730,1790  & 1.5 & Effelsberg & 1205\\
024503.71+004322.3  & 0.315 &  8.22  &	       & 16908.81 & 21.4 & --1660,1600  & 1.4 & Effelsberg & 1205\\
024545.44+002513.7  & 0.361 &  7.88  &	       & 16337.31 & 12.0 & --1650,1720  & 1.4 & Effelsberg & 1205\\
024607.92--000532.0 & 0.493 &  8.26  &	       & 14892.90 & 1.4  & --2010,2010  & 0.5 & GBT 	   & 0605\\
024919.70+010042.8  & 0.370 &  7.86  &	       & 14081.70 & 2.2  & --2125,2125  & 0.5 & GBT 	   & 0605\\
024946.09+001003.1  & 0.408 &  8.63  &  1.03   & 15791.96 & 11.9 & --1720,1780  & 1.5 & Effelsberg & 1205\\
025133.72--001146.2 & 0.318 &  8.10  &	       & 16870.32 & 24.6 & --1680,1610  & 1.4 & Effelsberg & 1205\\
025134.56+001308.9  & 0.346 &  7.60  &	       & 16519.38 & 13.9 & --1650,1700  & 1.4 & Effelsberg & 1205\\
025558.00--005954.0 & 0.700 &  8.51  &	       & 13079.50 & 2.0  & --2290,2290  & 0.5 & GBT 	   & 0105\\
                    &       &        &         & 13079.50 & 2.2  & --2290,2290  & 0.6 & GBT 	   & 0605\\
025725.99--063205.4 & 0.557 &  8.20  & 20.89   & 14280.70 & 2.1  & --2095,2095  & 0.5 & GBT 	   & 0605\\
025951.28+002301.0  & 0.505 &  8.53  &	       & 14774.10 & 1.8  & --2025,2025  & 0.5 & GBT 	   & 0105\\
030545.47--010010.5 & 0.359 &  8.17  &  0.80   & 16361.35 & 10.4 & --1670,1720  & 1.4 & Effelsberg & 1205\\
030809.79+005225.8  & 0.466 &  8.43  &	       & 15167.20 & 2.1  & --1975,1975  & 0.5 & GBT 	   & 0105\\
031012.82--010822.6 & 0.303 &  8.06  &	       & 17064.53 & 18.6 & --1640,1600  & 1.4 & Effelsberg & 1205\\
031319.96+003715.6  & 0.395 &  7.93  &	       & 15939.12 & 8.9  & --1700,1760  & 1.5 & Effelsberg & 1205\\
031449.11--010502.3 & 0.557 &  9.14  &	       & 14280.70 & 2.0  & --2095,2095  & 0.5 & GBT 	   & 0105\\
                    &       &        &         & 14280.70 & 1.9  & --2095,2095  & 0.5 & GBT 	   & 0605\\
031606.01+004733.1  & 0.370 &  7.90  &	       & 16229.99 & 12.2 & --1670,1740  & 1.4 & Effelsberg & 1205\\
031636.15--005634.0 & 0.470 &  8.54  &	       & 15125.90 & 2.4  & --1980,1980  & 0.5 & GBT 	   & 0105\\
031643.56--004343.2 & 0.379 &  7.71  &	       & 16124.06 & 11.6 & --1710,1750  & 1.5 & Effelsberg & 1205\\
031645.60--005931.0 & 0.369 &  8.33  &  7.79   & 16241.84 & 12.7 & --1680,1720  & 1.4 & Effelsberg & 1205\\
031927.22+000014.5  & 0.385 &  8.06  &	       & 16054.00 & 11.4 & --1680,1750  & 1.5 & Effelsberg & 1205\\
031946.03--001629.1 & 0.393 &  8.24  &	       & 15962.01 & 10.3 & --1700,1760  & 1.5 & Effelsberg & 1205\\
031947.27--010504.0 & 0.699 &  8.44  &	       & 13087.20 & 2.0  & --2290,2290  & 0.5 & GBT 	   & 0605\\
031950.54--005850.6 & 0.626 &  9.58  &	       & 13674.70 & 2.2  & --2190,2190  & 0.5 & GBT 	   & 0105\\
                    &       &        &         & 13674.70 & 5.0  & --2190,2190  & 0.5 & GBT 	   & 0605\\
032029.78+003153.5  & 0.384 &  8.52  &	       & 16065.81 & 11.8 & --1690,1740  & 1.5 & Effelsberg & 1205\\
032107.80+005901.4  & 0.529 &  8.05  & n/a     & 14542.20 & 2.2  & --2060,2060  & 0.5 & GBT 	   & 0605\\
032224.93+002004.3  & 0.346 &  7.49  & n/a     & 16519.38 & 12.9 & --1630,1700  & 1.4 & Effelsberg & 1205\\
032240.60+001626.0  & 0.344 &  7.58  & n/a     & 16543.96 & 14.1 & --1640,1700  & 1.4 & Effelsberg & 1205\\
032939.85+005220.0  & 0.446 &  8.28  & n/a     & 15377.00 & 2.4  & --1945,1945  & 0.5 & GBT 	   & 0105\\
033216.38+002618.2  & 0.353 &  7.79  & n/a     & 16433.91 & 12.6 & --1660,1700  & 1.4 & Effelsberg & 1205\\
033248.50--001012.3 & 0.310 &  8.50  & n/a     & 16973.34 & 24.3 & --1660,1590  & 1.4 & Effelsberg & 1205\\
033310.10+000849.1  & 0.327 &  8.13  & n/a     & 16755.90 & 27.7 & --1680,1610  & 1.4 & Effelsberg & 1205\\
033435.47+003724.9  & 0.407 &  8.61  & n/a     & 15803.18 & 7.4  & --500,1310   & 1.5 & Effelsberg & 1105\\
033606.70--000754.7 & 0.431 &  8.71  & n/a     & 15538.20 & 1.0  & --1925,1925  & 0.5 & GBT 	   & 0605\\
034252.47+005252.4  & 0.565 &  8.74  & n/a     & 14207.70 & 2.3  & --2110,2110  & 0.5 & GBT 	   & 0605\\
034416.79--010105.4 & 0.306 &  7.57  & n/a     & 17025.33 & 24.1 & --1650,1600  & 1.4 & Effelsberg & 1205\\
073705.07+324033.3  & 0.532 &  8.65  & 13.87   & 14513.70 & 2.0  & --2065,2065  & 0.5 & GBT 	   & 0105\\
073745.88+402146.5  & 0.613 &  9.28  &	       & 13784.90 & 2.0  & --2170,2170  & 0.5 & GBT 	   & 0605\\
073910.48+333353.8  & 0.446 &  9.11  &  0.81   & 15377.00 & 2.0  & --1945,1945  & 0.5 & GBT 	   & 0105\\
074130.51+302005.3  & 0.476 &  8.33  &  1.08   & 15064.40 & 1.9  & --1990,1990  & 0.5 & GBT 	   & 0605\\
074254.90+344236.5  & 0.567 &  8.32  &  1.67   & 14189.60 & 1.9  & --2110,2110  & 0.5 & GBT 	   & 0105\\
074430.86+394505.1  & 0.486 &  8.59  &  0.69   & 14963.00 & 1.9  & --2000,2000  & 0.5 & GBT 	   & 0605\\
074520.32+391839.9  & 0.339 &  7.78  &	       & 16605.74 & 12.5 & --1630,1700  & 1.4 & Effelsberg & 1205\\
074811.44+395238.0  & 0.372 &  8.19  &  2.83   & 16206.33 & 11.5 & --1680,1750  & 1.4 & Effelsberg & 1205\\
074900.11+351823.0  & 0.466 &  7.59  &  4.24   & 15167.20 & 1.9  & --1975,1975  & 0.5 & GBT 	   & 0105\\
075129.33+403211.1  & 0.359 &  7.25  &  1.49   & 16361.35 & 9.9  & --1650,1720  & 1.4 & Effelsberg & 1205\\
075238.68+390304.9  & 0.654 &  8.49  &  2.96   & 13443.20 & 2.1  & --2230,2229  & 0.5 & GBT 	   & 0605\\
075607.16+461411.5  & 0.593 &  9.01  & 20.50   & 13958.00 & 1.9  & --2145,2145  & 0.5 & GBT 	   & 0105\\
                    &	    &	     &         & 13958.00 & 1.9  & --2145,2145  & 0.5 & GBT 	   & 0605\\
075920.21+351903.4  & 0.328 &  7.59  &         & 16743.28 & 20.7 & --1680,1630  & 1.4 & Effelsberg & 1205\\
080154.24+441234.0  & 0.556 &  9.58  &         & 14289.90 & 1.8  & --2095,2095  & 0.5 & GBT 	   & 0105\\
                    &	    &	     &         & 14289.90 & 2.0  & --2095,2095  & 0.5 & GBT 	   & 0605\\
080338.58+412045.5  & 0.503 &  8.52  &  0.99   & 14793.80 & 2.1  & --2025,2025  & 0.5 & GBT 	   & 0605\\
081330.42+320506.0  & 0.398 &  8.83  &  2.33   & 15904.92 & 4.7  & --500,1300	& 1.5 & Effelsberg & 1105\\
081507.42+430427.2  & 0.510 &  9.57  &  6.09   & 14725.20 & 2.3  & --2035,2035  & 0.5 & GBT 	   & 0605\\
081836.46+460942.9  & 0.352 &  7.96  &  0.88   & 16446.07 & 11.4 & --1650,1700  & 1.4 & Effelsberg & 1205\\
081858.36+395839.8  & 0.406 &  8.23  &  5.77   & 15814.42 & 11.2 & --1700,1770  & 1.5 & Effelsberg & 1205\\
082449.27+370355.7  & 0.305 &  8.28  &  3.59   & 17038.38 & 19.6 & --1660,1600  & 1.4 & Effelsberg & 1205\\
083620.35+470357.3  & 0.423 &  8.42  &         & 15625.49 & 12.0 & --1720,1800  & 1.5 & Effelsberg & 1205\\
083754.60+391045.8  & 0.633 &  8.96  &  0.89   & 13616.10 & 2.2  & --2200,2200  & 0.5 & GBT 	   & 0605\\
083945.98+384319.0  & 0.424 &  8.70  &  1.67   & 15614.52 & 6.9  & --510,1320	& 1.5 & Effelsberg & 1105\\
084015.49+504342.4  & 0.402 &  8.56  &  6.80   & 15859.54 & 5.6  & --500,1300	& 1.5 & Effelsberg & 1105\\
084041.08+383819.8  & 0.313 &  8.62  &  1.16   & 16934.56 & 9.5  & --510,1140	& 1.4 & Effelsberg & 1105\\
084148.93+384602.1  & 0.492 &  7.99  &  1.14   & 14902.90 & 2.0  & --2010,2010  & 0.5 & GBT 	   & 0605\\
084846.37+022034.1  & 0.627 &  8.59  &  0.67   & 13666.30 & 3.3  & --2190,2190  & 0.5 & GBT 	   & 0605\\
084856.58+013647.8  & 0.350 &  8.56  & 88.24   & 16470.43 & 6.3  & --480,1220	& 1.4 & Effelsberg & 1105\\
084943.82+015058.2  & 0.376 &  8.06  &  1.14   & 16159.22 & 11.4 & --1700,1740  & 1.4 & Effelsberg & 1205\\
085049.76+462627.1  & 0.554 &  8.97  &  3.38   & 14308.30 & 1.8  & --2095,2095  & 0.5 & GBT 	   & 0105\\
                    &	    &	     &         & 14308.29 & 7.6  & --520,1450	& 1.6 & Effelsberg & 1105\\
085305.94+005006.1  & 0.433 &  8.03  &  4.28   & 15516.50 & 2.1  & --1930,1930  & 0.5 & GBT 	   & 0605\\
085314.23+021453.9  & 0.460 &  7.85  & 119.74  & 15229.50 & 2.1  & --1965,1965  & 0.5 & GBT 	   & 0605\\
085512.14+483609.1  & 0.554 &  8.89  &         & 14308.30 & 2.1  & --2095,2095  & 0.5 & GBT 	   & 0605\\
085704.96--001707.3 & 0.329 &  7.83  &  1.47   & 16730.68 & 20.7 & --1690,1600  & 1.4 & Effelsberg & 1205\\
085852.06+470150.2  & 0.446 &  7.92  &         & 15377.00 & 2.1  & --1945,1945  & 0.5 & GBT 	   & 0605\\
                    &	    &	     &         & 15376.96 & 13.4 & --1750,1820  & 1.5 & Effelsberg & 1205\\
090226.74+545952.3  & 0.401 &  8.43  &         & 15870.86 & 10.1 & --1720,1760  & 1.5 & Effelsberg & 1205\\
090246.94+012028.4  & 0.513 &  9.20  &  1.38   & 14696.00 & 1.7  & --2035,2035  & 0.5 & GBT 	   & 0105\\
090249.97+031341.1  & 0.457 &  8.14  &  2.36   & 15260.90 & 2.0  & --1960,1960  & 0.5 & GBT 	   & 0605\\
090307.84+021152.2  & 0.329 &  8.42  & 22.50   & 16730.68 & 20.7 & --1680,1610  & 1.4 & Effelsberg & 1205\\
090320.46+523336.1  & 0.315 &  7.89  &         & 16908.81 & 20.2 & --1780,1590  & 1.4 & Effelsberg & 1205\\
090414.10--002144.9 & 0.353 &  8.93  &  2.34   & 16433.91 & 5.8  & --490,1250   & 1.4 & Effelsberg & 1105\\
090626.80+033310.7  & 0.363 &  7.73  &  2.68   & 16313.34 & 11.1 & --1680,1720  & 1.4 & Effelsberg & 1205\\
090801.32+434722.6  & 0.363 &  8.31  & 32.05   & 16313.34 & 11.5 & --1670,1720  & 1.4 & Effelsberg & 1205\\
090933.51+425346.5  & 0.670 &  8.92  & 4009.50 & 13314.40 & 2.4  & --2250,2250  & 0.5 & GBT 	   & 0105\\
091157.55+014327.6  & 0.603 &  8.80  &  4.56   & 13870.90 & 2.2  & --2160,2160  & 0.5 & GBT 	   & 0605\\
                    &	    &	     &         & 13870.92 & 12.2 & --1960,2020  & 1.7 & Effelsberg & 1205\\
091231.97+451420.6  & 0.306 &  7.91  &  3.61   & 17025.33 & 21.8 & --1660,1580  & 1.4 & Effelsberg & 1205\\
091442.33+000637.2  & 0.561 &  9.13  &  1.44   & 14244.10 & 1.8  & --2100,2100  & 0.5 & GBT 	   & 0105\\
092014.11+453157.3  & 0.402 &  9.04  &         & 15859.54 & 5.1  & --490,1300	& 1.5 & Effelsberg & 1105\\
                    &	    &	     &         & 15859.54 & 9.8  & --1700,1770  & 1.5 & Effelsberg & 1205\\
092152.45+515348.1  & 0.587 &  9.28  &  2.37   & 14010.80 & 1.8  & --2135,2135  & 0.5 & GBT 	   & 0105\\
092223.65+020914.8  & 0.546 &  8.51  &  3.36   & 14382.30 & 2.1  & --2080,2080  & 0.5 & GBT 	   & 0605\\
092318.06+010144.8  & 0.386 &  8.94  &  1.11   & 16042.63 & 11.3 & --1700,1740  & 1.5 & Effelsberg & 1205\\
092356.44+012002.1  & 0.380 &  8.59  & 14.74   & 16112.38 & 6.2  & --520,1280	& 1.5 & Effelsberg & 1105\\
093818.57+005826.8  & 0.493 &  8.09  &  0.89   & 14892.90 & 1.4  & --2010,2010  & 0.5 & GBT 	   & 0605\\
094209.00+570019.7  & 0.350 &  8.31  &  0.93   & 16470.43 & 14.6 & --1660,1700  & 1.4 & Effelsberg & 1205\\
094312.82+024325.8  & 0.592 &  9.14  & 906.39  & 13966.80 & 1.8  & --2145,2145  & 0.5 & GBT 	   & 0105\\
094330.68+562530.8  & 0.347 &  7.67  &         & 16507.11 & 9.0  & --1660,1700  & 1.4 & Effelsberg & 1205\\
094350.92+610255.9  & 0.341 &  8.46  &         & 16580.97 & 13.7 & --1610,1700  & 1.4 & Effelsberg & 1205\\
094557.03+570803.2  & 0.512 &  8.32  & 25.19   & 14705.70 & 1.8  & --2035,2035  & 0.5 & GBT 	   & 0105\\
094820.38+582526.6  & 0.353 &  7.89  &         & 16433.91 & 11.4 & --1660,1710  & 1.4 & Effelsberg & 1205\\
094836.05+002104.6  & 0.324 &  8.52  &  2.21   & 16793.87 & 17.8 & --520,1120	& 1.4 & Effelsberg & 1105\\
095044.69+011127.2  & 0.404 &  8.22  &  1.68   & 15836.95 & 11.2 & --1720,1770  & 1.5 & Effelsberg & 1205\\
095126.49+014651.8  & 0.494 &  8.55  & 10.52   & 14882.90 & 1.7  & --2010,2010  & 0.5 & GBT 	   & 0105\\
095514.11+034654.2  & 0.421 &  8.60  &  1.93   & 15647.49 & 12.0 & --1720,1800  & 1.5 & Effelsberg & 1205\\
095629.06+573508.9  & 0.361 &  8.38  &  1.13   & 16337.31 & 11.3 & --1650,1720  & 1.4 & Effelsberg & 1205\\
095906.61+510325.3  & 0.570 &  8.95  &         & 14162.50 & 2.0  & --2115,2115  & 0.5 & GBT 	   & 0105\\
095941.73+580545.9  & 0.465 &  8.21  &         & 15177.50 & 2.3  & --1975,1975  & 0.5 & GBT 	   & 0605\\
100329.86+511630.7  & 0.324 &  8.11  &  2.74   & 16793.87 & 27.7 & --1680,1610  & 1.4 & Effelsberg & 1205\\
100459.41+030202.0  & 0.469 &  8.77  &         & 15136.20 & 1.8  & --1980,1980  & 0.5 & GBT 	   & 0105\\
100854.43+461300.7  & 0.544 &  8.32  &  7.16   & 14401.00 & 1.9  & --2080,2080  & 0.5 & GBT 	   & 0605\\
101005.88+010703.0  & 0.334 &  7.95  &  1.44   & 16667.98 & 11.8 & --1680,1620  & 1.4 & Effelsberg & 1205\\
101104.45+010333.4  & 0.306 &  8.11  &         & 17025.33 & 19.6 & --1660,1600  & 1.4 & Effelsberg & 1205\\
101120.22+444225.7  & 0.558 &  8.98  &  1.75   & 14271.60 & 1.9  & --2100,2100  & 0.5 & GBT 	   & 0105\\
101237.32+023554.3  & 0.720 &  8.22  &  1.03   & 12927.40 & 2.4  & --2315,2315  & 0.6 & GBT 	   & 0605\\
102640.42--004206.5 & 0.365 &  8.33  &  1.94   & 16298.44 & 11.8 & --1650,1720  & 1.4 & Effelsberg & 1205\\
102746.03+003205.0  & 0.614 &  9.36  &  3.25   & 13776.40 & 2.0  & --2175,2175  & 0.5 & GBT 	   & 0105\\
102856.00+571223.8  & 0.437 &  8.93  &  3.05   & 15473.30 & 2.1  & --1935,1935  & 0.5 & GBT 	   & 0605\\
103622.19+635553.0  & 0.432 &  8.62  &         & 15527.30 & 2.2  & --1930,1930  & 0.5 & GBT 	   & 0605\\
103639.39+640924.7  & 0.398 &  8.42  & n/a     & 15904.92 & 4.7  & --500,1290	& 1.5 & Effelsberg & 1105\\
103951.49+643004.2  & 0.402 &  9.41  & n/a     & 15859.54 & 5.6  & --510,1300	& 1.5 & Effelsberg & 1105\\
104210.95+001048.3  & 0.397 &  8.04  &         & 15916.31 & 10.3 & --1700,1760  & 1.5 & Effelsberg & 1205\\
104505.39+561118.4  & 0.428 &  9.08  &  2.59   & 15570.80 & 2.2  & --1925,1925  & 0.5 & GBT 	   & 0605\\
104807.74+005543.4  & 0.642 &  9.37  & 270.70  & 13541.50 & 2.9  & --2210,2210  & 0.5 & GBT 	   & 0105\\
105809.03+610527.1  & 0.343 &  7.88  &         & 16556.28 & 12.7 & --1630,1700  & 1.4 & Effelsberg & 1205\\
110321.85+600520.6  & 0.669 &  8.69  &         & 13322.40 & 2.0  & --2250,2249  & 0.5 & GBT 	   & 0105\\
110709.36+511328.6  & 0.441 &  8.09  &         & 15430.30 & 2.3  & --1940,1940  & 0.5 & GBT 	   & 0605\\
111112.87+030850.3  & 0.461 &  8.30  &         & 15219.10 & 2.5  & --1965,1965  & 0.5 & GBT 	   & 0605\\
113344.02+613455.7  & 0.426 &  8.82  &         & 15592.60 & 2.0  & --1920,1920  & 0.5 & GBT 	   & 0605\\
115314.36+032658.6  & 0.575 &  9.61  &  1.62   & 14117.50 & 1.8  & --2120,2120  & 0.5 & GBT 	   & 0105\\
115718.35+600345.6  & 0.491 &  9.60  &         & 14912.90 & 2.0  & --2010,2010  & 0.5 & GBT 	   & 0105\\
115954.43--012108.3 & 0.374 &  7.90  &	       & 16182.74 & 12.2 & --1680,1710  & 1.4 & Effelsberg & 1205\\
121541.79+005737.7  & 0.432 &  7.72  &         & 15527.30 & 2.1  & --1930,1930  & 0.5 & GBT 	   & 0605\\
121637.27+672441.6  & 0.362 &  8.18  & n/a     & 16325.32 & 8.7  & --490,1260	& 1.4 & Effelsberg & 1105\\
121856.42+611922.7  & 0.369 &  8.38  &         & 16241.84 & 8.6  & --500,1270	& 1.4 & Effelsberg & 1105\\
122408.45--022312.6 & 0.434 &  7.64  & 33.03   & 15505.60 & 2.3  & --1930,1930  & 0.5 & GBT 	   & 0605\\
122656.48+013124.3  & 0.732 &  9.66  &         & 12837.80 & 2.1  & --2335,2335  & 0.5 & GBT 	   & 0105\\
122845.74+005018.7  & 0.575 &  9.28  &  3.14   & 14117.50 & 2.0  & --2120,2120  & 0.5 & GBT 	   & 0105\\
                    &	    &	     &         & 14117.50 & 1.6  & --2120,2120  & 0.5 & GBT 	   & 0605\\   
123115.48+514929.1  & 0.474 &  8.17  &  1.36   & 15084.90 & 2.0  & --1985,1985  & 0.5 & GBT 	   & 0605\\
123215.81+020610.0  & 0.480 &  9.69  &         & 15023.70 & 2.0  & --1995,1995  & 0.5 & GBT 	   & 0105\\
123453.10+640510.2  & 0.594 &  8.77  &         & 13949.20 & 2.2  & --2145,2145  & 0.5 & GBT 	   & 0105\\
124736.07+023110.7  & 0.487 &  8.59  &  0.90   & 14953.00 & 1.5  & --2000,2000  & 0.5 & GBT 	   & 0605\\
124749.79+015212.6  & 0.427 &  8.23  &  8.04   & 15581.70 & 2.0  & --1920,1920  & 0.5 & GBT 	   & 0605\\
124806.80+593957.6  & 0.453 &  8.92  &         & 15302.90 & 2.4  & --1955,1955  & 0.5 & GBT 	   & 0105\\
130740.56--021455.3 & 0.425 &  8.92  &  1.66   & 15603.56 & 6.9  & --480,1300   & 1.5 & Effelsberg & 1105\\
132323.33--015941.9 & 0.350 &  9.19  &  0.74   & 16470.40 & 9.7  & --480,1250   & 1.4 & Effelsberg & 1105\\
                    &       &        &         & 16470.40 & 10.0 & --1670,1710  & 1.4 & Effelsberg & 1205\\
132529.33+592424.9  & 0.429 &  8.89  &	       & 15559.90 & 2.0  & --1925,1925  & 0.5 & GBT 	   & 0605\\
133550.36--012439.3 & 0.523 &  9.02  &  3.21   & 14599.50 & 1.9  & --2050,2050  & 0.5 & GBT 	   & 0605\\
133633.65--003936.4 & 0.416 &  8.64  &         & 15702.74 & 6.9  & --500,1290   & 1.5 & Effelsberg & 1105\\
133735.02--012815.7 & 0.329 &  8.72  &  1.58   & 16730.68 & 8.9  & --530,1180   & 1.4 & Effelsberg & 1105\\
133747.57--024247.8 & 0.499 &  7.95  &  0.95   & 14833.30 & 1.0  & --2020,2020  & 0.5 & GBT 	   & 0605\\
135128.14--001016.9 & 0.524 &  8.47  &  2.02   & 14590.00 & 2.7  & --2050,2050  & 0.5 & GBT 	   & 0605\\
140550.31+641947.1  & 0.332 &  7.83  & n/a     & 16693.00 & 17.2 & --530,1170   & 1.4 & Effelsberg & 1105\\
140740.06+021748.3  & 0.309 &  8.90  &  1.07   & 16986.31 & 10.0 & --520,1140   & 1.4 & Effelsberg & 1105\\
141315.31--014221.0 & 0.380 &  9.25  &  3.07   & 16112.38 & 5.7  & --500,1290   & 1.5 & Effelsberg & 1105\\
141611.77--023117.1 & 0.305 &  8.03  &  1.37   & 17038.38 & 22.6 & --1650,1600  & 1.4 & Effelsberg & 1205\\
141855.62--015558.3 & 0.372 &  7.76  &  2.25   & 16206.33 & 17.0 & --1660,1730  & 1.4 & Effelsberg & 1205\\
142124.08--003354.6 & 0.454 &  7.94  &	       & 15292.30 & 2.4  & --1960,1960  & 0.5 & GBT 	   & 0605\\
143027.66--005614.9 & 0.318 &  8.36  &	       & 16870.32 & 25.9 & --1670,1600  & 1.4 & Effelsberg & 1205\\
143047.33+602304.5  & 0.607 &  8.44  &         & 13836.40 & 2.1  & --2165,2165  & 0.5 & GBT 	   & 0605\\
143233.70+030946.7  & 0.365 &  8.01  &  2.14   & 16289.44 & 13.0 & --1680,1720  & 1.4 & Effelsberg & 1205\\
143731.86+011858.2  & 0.342 &  8.11  & 114.86  & 16568.61 & 14.7 & --1630,1700  & 1.4 & Effelsberg & 1205\\
143928.24+001537.9  & 0.339 &  8.08  &  1.07   & 16605.74 & 15.4 & --1630,1700  & 1.4 & Effelsberg & 1205\\
144642.29+011303.0  & 0.725 &  9.37  &  5.75   & 12889.90 & 2.3  & --2325,2325  & 0.6 & GBT 	   & 0605\\
144711.29+021136.2  & 0.386 &  8.45  &  0.70   & 16043.00 & 11.3 & --1690,1750  & 1.5 & Effelsberg & 1205\\
144943.54+045154.9  & 0.450 &  7.90  &         & 15334.60 & 1.7  & --1955,1955  & 0.5 & GBT 	   & 0605\\
145054.37+004646.7  & 0.335 &  7.72  &  1.29   & 16655.49 & 16.9 & --1690,1620  & 1.4 & Effelsberg & 1205\\
145201.73+005040.2  & 0.314 &  7.82  &  0.53   & 16921.67 & 20.5 & --1660,1580  & 1.4 & Effelsberg & 1205\\
150043.58+625700.5  & 0.303 &  7.93  &         & 17064.53 & 11.6 & --520,1120	& 1.4 & Effelsberg & 1105\\
150117.96+545518.3  & 0.338 &  9.06  & 20.87   & 16618.15 & 16.2 & --1630,1690  & 1.4 & Effelsberg & 1205\\
150608.09--020744.2 & 0.439 &  9.25  &	       & 15451.80 & 2.2  & --1940,1940  & 0.5 & GBT 	   & 0605\\
150919.06+542606.4  & 0.329 &  7.62  &  1.45   & 16370.68 & 23.4 & --1680,1620  & 1.4 & Effelsberg & 1205\\
151325.10+534527.2  & 0.560 &  8.90  &  1.29   & 14253.20 & 2.6  & --2100,2100  & 0.5 & GBT 	   & 0105\\
151711.47+033100.2  & 0.613 &  9.36  &	       & 13784.90 & 2.4  & --2170,2170  & 0.5 & GBT 	   & 0605\\
152019.75--013611.2 & 0.307 &  8.29  &  1.40   & 17012.30 & 20.6 & --1650,1610  & 1.4 & Effelsberg & 1205\\
153446.75+542008.2  & 0.364 &  7.90  &  2.36   & 16301.38 & 6.2  & --500,1270   & 1.4 & Effelsberg & 1105\\
153734.00+511258.9  & 0.444 &  8.30  &  0.92   & 15398.20 & 2.6  & --1945,1945  & 0.5 & GBT 	   & 0105\\
153943.73+514221.0  & 0.585 &  8.47  & 14.46   & 14028.50 & 1.9  & --2135,2135  & 0.5 & GBT 	   & 0605\\
154133.19+521200.1  & 0.305 &  8.25  &	       & 17038.38 & 24.1 & --1650,1600  & 1.4 & Effelsberg & 1205\\
154337.82--004419.9 & 0.311 &  8.40  &  1.62   & 16960.40 & 22.3 & --1650,1600  & 1.4 & Effelsberg & 1205\\
154340.02+493512.6  & 0.512 &  9.13  &	       & 14705.70 & 2.5  & --2035,2035  & 0.5 & GBT 	   & 0105\\
154613.27--000513.5 & 0.383 &  8.18  &  1.66   & 16077.43 & 7.1  & --480,1260   & 1.5 & Effelsberg & 1105\\
154826.05+004615.3  & 0.544 &  8.33  &  4.24   & 14401.00 & 2.0  & --2080,2080  & 0.5 & GBT 	   & 0605\\
160448.83+001550.5  & 0.337 &  7.79  &         & 16630.58 & 7.8  & --490,1220	& 1.4 & Effelsberg & 1105\\
164019.72+385637.5  & 0.418 &  8.63  &  1.07   & 15680.59 & 6.0  & --500,1320	& 1.5 & Effelsberg & 1105\\
164131.73+385840.9  & 0.596 &  9.92  &  2.80   & 13931.80 & 2.0  & --2150,2150  & 0.5 & GBT 	   & 0605\\
164351.40+375721.3  & 0.582 &  8.72  &         & 14055.10 & 2.0  & --2130,2130  & 0.5 & GBT 	   & 0605\\
165627.28+351401.7  & 0.679 &  8.57  &  1.16   & 13243.10 & 1.6  & --2260,2260  & 0.5 & GBT 	   & 0605\\
170151.98+385901.4  & 0.596 &  9.09  &         & 13931.80 & 2.0  & --2150,2150  & 0.5 & GBT 	   & 0605\\
171559.79+280716.8  & 0.524 &  9.17  &  0.94   & 14590.00 & 1.5  & --2050,2050  & 0.5 & GBT 	   & 0605\\ 
171629.37+301416.5  & 0.486 &  8.51  & 24.48   & 14963.00 & 1.5  & --2000,2000  & 0.5 & GBT 	   & 0605\\
172419.89+551058.8  & 0.365 &  8.00  &         & 16289.44 & 6.2  & --500,1280	& 1.4 & Effelsberg & 1105\\
172603.09+602115.7  & 0.333 &  8.57  &         & 16680.48 & 7.9  & --540,1180	& 1.4 & Effelsberg & 1105\\
173938.64+544208.6  & 0.384 &  8.42  &  1.14   & 16065.81 & 7.6  & --500,1290	& 1.5 & Effelsberg & 1105\\
211742.59+005708.0  & 0.486 &  8.78  & n/a     & 14963.00 & 2.5  & --2000,2000  & 0.5 & GBT 	   & 0105\\
214415.61+125503.0  & 0.390 &  8.14  & n/a     & 15996.46 & 5.2  & --520,1290	& 1.5 & Effelsberg & 1105\\
214800.73--002834.1 & 0.515 &  8.58  & 32.62   & 14676.60 & 2.0  & --2040,2040  & 0.5 & GBT 	   & 0105\\
215731.40+003757.1  & 0.390 &  8.39  &         & 15996.46 & 7.5  & --500,1300   & 1.5 & Effelsberg & 1105\\
222310.44--004330.5 & 0.493 &  8.59  &	       & 14892.90 & 2.3  & --2010,2010  & 0.5 & GBT 	   & 0105\\
222631.14--010054.0 & 0.530 &  8.46  &	       & 14532.70 & 1.8  & --2060,2060  & 0.5 & GBT 	   & 0105\\
223136.27--011045.0 & 0.436 &  8.60  &	       & 15484.10 & 2.1  & --1935,1935  & 0.5 & GBT 	   & 0605\\
223841.50--094404.0 & 0.433 &  8.59  &	       & 15516.50 & 2.3  & --1930,1930  & 0.5 & GBT 	   & 0605\\
223959.04+005138.3  & 0.384 &  8.15  &         & 16065.81 & 5.7  & --490,1260   & 1.5 & Effelsberg & 1105\\
224409.48--083505.2 & 0.617 &  8.64  &	       & 13750.80 & 2.3  & --2180,2180  & 0.5 & GBT 	   & 0605\\
224950.42--005157.2 & 0.597 &  8.84  &	       & 13923.00 & 1.1  & --2150,2150  & 0.5 & GBT 	   & 0105\\
                    &       &        &         & 13923.00 & 2.5  & --2153,2152  & 0.5 & GBT 	   & 0605\\
225102.40--000459.9 & 0.550 &  9.13  &  1.67   & 14345.20 & 1.9  & --2085,2085  & 0.5 & GBT 	   & 0105\\
                    &       &        &         & 14345.21 & 5.5  & --540,1400   & 1.6 & Effelsberg & 1105\\
225227.39--005528.5 & 0.442 &  8.23  &  1.06   & 15419.60 & 2.4  & --1940,1940  & 0.5 & GBT 	   & 0605\\
225612.18--010508.1 & 0.648 &  9.12  &  2.98   & 13492.10 & 2.0  & --2220,2220  & 0.5 & GBT 	   & 0105\\
                    &       &        &         & 13492.20 & 2.8  & --2220,2220  & 0.5 & GBT 	   & 0605\\
225721.78--100000.9 & 0.426 &  8.12  &	       & 15592.60 & 2.3  & --1920,1920  & 0.5 & GBT 	   & 0605\\
230307.46+004632.7  & 0.378 &  7.75  &  3.34   & 16135.76 & 7.6  & --500,1280   & 1.5 & Effelsberg & 1105\\
230937.14+001735.8  & 0.555 &  8.34  &	       & 14299.10 & 2.0  & --2095,2095  & 0.5 & GBT 	   & 0605\\
231239.40--005005.5 & 0.479 &  9.07  &	       & 15033.80 & 1.9  & --1990,1990  & 0.5 & GBT 	   & 0105\\
231606.89--002519.1 & 0.497 &  8.76  &	       & 14853.10 & 2.9  & --2015,2015  & 0.5 & GBT 	   & 0605\\
231755.35+145349.4  & 0.311 &  8.10  &  n/a    & 16960.40 & 9.5  & --1660,1600  & 1.4 & Effelsberg & 1205\\
231845.12--002951.4 & 0.397 &  8.00  &         & 15916.31 & 4.7  & --500,1300   & 1.5 & Effelsberg & 1105\\
232902.94--002717.0 & 0.620 &  8.73  &	       & 13725.40 & 2.1  & --2180,2180  & 0.5 & GBT 	   & 0605\\
232942.17+000521.2  & 0.446 &  7.89  &  0.97   & 15377.00 & 2.0  & --1945,1945  & 0.5 & GBT 	   & 0605\\
235433.86--005629.3 & 0.348 &  8.22  &  2.22   & 16494.87 & 6.8  & --490,1240   & 1.4 & Effelsberg & 1105\\
235818.87--000919.5 & 0.402 &  9.32  &         & 15859.54 & 6.1  & --500,1280   & 1.5 & Effelsberg & 1105\\
235831.16--002226.5 & 0.628 &  9.96  &	       & 13657.90 & 2.2  & --2195,2195  & 0.5 & GBT 	   & 0105\\
235844.52--010611.9 & 0.684 &  8.53  &	       & 13203.70 & 2.1  & --2270,2270  & 0.5 & GBT  	   & 0605\\
\enddata
\tablecomments{
Col. (1): Source; bold-face: the only object for which
megamaser emission was detected \citep{bar05}. Col. (2): Heliocentric redshift from \citet{zak03}
as measured from the [OII]\,$\lambda$3727 emission line. Col. (3): 
Log ($L_{\rm [OIII]}/L_{\odot}$) taken from \citet{zak03}.
A value of log ($L_{\rm [OIII]}/L_{\odot}$) $>$ 8.48 classifies the object as
a type-2 QSO. 
Col. (4): FIRST integrated fluxes at 20 cm in mJy taken from \citet{zak03}, if the
object is matched within 3\arcsec. There is no entry if the object was not detected
[$F_\nu$ (20 cm) $<$ 1 mJy]. ``n/a'' denotes those objects for which the field
was not observed by the FIRST survey.
Col. (5): Observed Frequency $\nu$ in MHz. 
Col. (6): Root mean square (rms) flux density in mJy.
Col. (7): Velocity range covered by observations in km\,s$^{-1}$.
Col. (8): Channel width in km\,s$^{-1}$.
Col. (9): Telescope at which the source was observed.
Col. (10): Date of observations (mmyy).
}
\label{table1}
\end{deluxetable}

\begin{deluxetable}{lcccccccccc}
\rotate
\tabletypesize{\scriptsize}
\tablecolumns{11}
\tablewidth{0pc}
\tablecaption{The 78 galaxies at $D \ge 100$\,kpc with known H$_2$O masers.}
\tablehead{
\colhead{Source} & \colhead{RA} & \colhead{DEC} & \colhead{$V_{\rm syst}$}
& \colhead{$D$} & \colhead{log $L_{\rm FIR}$} & \colhead{$T_{\rm dust}$}
& \colhead{log $L_{\rm H_2O}$} & \colhead{Morph.} & \colhead{Type} & \colhead{Ref.}\\
& \multicolumn{2}{c}{(J2000)} & (km\,s$^{-1}$) & (Mpc) & ($L_{\odot}$) 
& (K) & ($L_{\odot}$) & & &\\
\colhead{(1)} & \colhead{(2)} & \colhead{(3)}  & \colhead{(4)} & \colhead{(5)}
& \colhead{(6)} & \colhead{(7)} & \colhead{(8)} & \colhead{(9)} & \colhead{(10)} & \colhead{(11)}}
\startdata
NGC~23 	            & 00 09 53.6 & +25 55 23 &  4566  &  60.9  & 11.1 	 & 45	   & 2.2   & SB(s)a        & L,LIRG,HII     & 1\\
IC~10               & 00 20 27.0 & +59 17 29 & --350  &  1.2   & 8.2	 & 40	   & --0.8 & dIRR          &		    & 2,3\\ 
                    & 00 20 17.9 & +59 18 31 &        &        &	 &  	   & --1.7 &               &		    & 4\\ 
NGC~235A 	    & 00 42 52.8 &--23 32 28 &  6519  &  86.9  & 11.4  	 & 44	   & 2.0   & S0 pec        & Sy1	    & 5\\
NGC~253             & 00 47 33.1 &--25 17 17 &   240  &  3.0   & 10.3	 & 52	   & --0.8 & SAB(s)c       & Sy2,SB,HII     & 6,7\\ 
                    & 00 47 33.6 &--25 17 14 &        &        &	 &	   & --1.7 &               &                & 7\\ 
NGC~262 (Mrk~348)   & 00 48 47.1 & +31 57 25 & 4505   &  62.0  & 10.4	 & 54	   & 2.6   & SA(s)0/a:     & Sy2	    & 8,9\\ 
ESO~013--G012       & 01 07 00.9 &--80 18 24 & 5045   &  67.0  & 10.7	 & 32	   & 2.7   & Sa            &		    & 10\\ 
NGC~449 (Mrk~1)     & 01 16 07.2 & +33 05 22 & 4780   &  64.0  & 10.7	 & 55	   & 1.7   & (R')S?        & Sy2	    & 11\\ 
NGC~520 	    & 01 24 34.9 & +03 47 30 & 2281   &  30.4  & 10.9    & 48      & 0.3   & Pec           & SB,HII         & 12\\
NGC~598 (M~33)$^a$  & 01 33 16.5 & +30 52 53 & --180  &  0.7   & 9.0	 & 36	   & --0.5 & SA(s)cd       & HII	    & 13,14,15,16\\ 
                    & 01 33 28.3 & +30 31 43 &        &        &	 &	   & --1.5 &               &     	    & 15,16\\ 
                    & 01 34 00.2 & +30 40 47 &        &        &         &         & --2   &               &                & 16,17\\
NGC~591 (Mrk~1157)  & 01 33 31.2 & +35 40 06 & 4555   &  61.0  & 10.5	 & 46	   & 1.4   & (R')SB0/a     & Sy2	    & 18\\ 
NGC~613 	    & 01 34 18.2 &--29 25 07 & 1481   &  17.9  & 10.4 	 & 38	   & 1.3   & SB(rs)bc      & Sy,HII	    & 5,12\\
IC~0184 	    & 01 59 51.2 &--06 50 25 & 5287   &  70.5  & 9.9$^b$ & 40$^b$  & 1.4   & SB(r)a:       & Sy2,HII        & 5\\
NGC~1052            & 02 41 04.8 &--08 15 21 & 1470   &  17.0  & 9.2	 & 54	   & 2.1   & E4            & Sy2,L	    & 11,20\\ 
NGC~1068            & 02 42 40.7 &--00 00 48 & 1135   &  14.5  & 11.2	 & 54	   & 2.2   & (R)SA(rs)b    & Sy1,Sy2	    & 21,22\\ 
NGC~1106 	    & 02 50 40.5 & +41 40 17 & 4337   &  57.8  & 10.3 	 & 49	   & 0.9   & SA0+          & Sy2	    & 1\\
Mrk~1066            & 02 59 58.6 & +36 49 14 & 3600   &  48.0  & 10.9	 & 55	   & 1.5   & (R)SB(s)0+    & Sy2	    & 18,23\\ 
NGC~1386            & 03 36 46.4 &--36 00 02 & 870    &  17.0  & 9.8	 & 46	   & 2.1   & SB(s)0+       & Sy2	    & 24\\ 
IRAS~03355+0104	    & 03 38 10.4 & +01 14 18 & 11926  &  159.0 & 11.2  	 & 36	   &2.7$^c$& S0/a          & Sy2	    & 19\\
IC~342$^d$          & 03 46 46.3 & +68 05 46 & 40     &  2.0   & 9.0	 & 49	   & --2.0 & SAB(rs)cd     & Sy2,HII	    & 25\\ 
UGC~3193 	    & 04 52 52.7 & +03 03 24 & 4454   &  59.4  & 10.6 	 & 43	   & 2.4   & SB(rs)ab:     &		    & 1\\
UGC~3255            & 05 09 50.2 & +07 29 00 & 5675   &  75.0  & 10.6	 & 40	   & 1.2   & SBb?          & Sy2	    & 18\\ 
Mrk~3               & 06 15 36.3 & +71 02 15 & 4010   &  54.0  & 10.7	 & 69	   & 1.0   & S0:           & Sy2	    & 18\\ 
NGC~2146            & 06 18 36.6 & +78 21 28 & 900    &  14.5  & 10.9	 & 53	   &0.0$^e$& SB(s)ab pec   & HII	    & 26\\ 
                    & 06 18 38.6 & +78 21 24 &        &        &	 &	   &0.0$^e$&               &     	    & 26\\ 
VII~Zw~073 	    & 06 30 25.6 & +63 40 41 & 11899  & 158.7  & 11.2  	 & 51	   & 2.2   &               & Sy2	    & 5\\
NGC~2273 	    & 06 50 08.7 & +60 50 45 &  1840  &  24.5  & 10.1    & 47	   & 0.8   & SB(r)a        & Sy2	    & 27\\
UGC~3789 	    & 07 19 30.9 & +59 21 18 &  3325  &  44.3  & 10.2    & 42	   & 2.5   & (R)SA(r)ab    &		    & 1\\
Mrk~78              & 07 42 41.7 & +65 10 37 & 11195  &  150.0 & 11.0	 & 60	   & 1.5   & SB            & Sy2	    & 18\\ 
Mrk~1210            & 08 04 05.8 & +05 06 50 & 4045   &  54.0  & 10.5	 & 75	   & 1.9   & Sa            & Sy1,Sy2	    & 11\\ 
SDSS~J0804+3607     & 08 04 31.0 & +36 07 18 & z=0.66 &3749.7$^f$&\nodata&\nodata  &4.3$^f$&               & QSO2	    & 28\\
2MASX~J0836+3327    & 08 36 22.8 & +33 27 39 & 14810  & 197.5  & \nodata &\nodata  &3.4$^c$&               & Sy2	    & 19\\
NGC~2639            & 08 43 38.1 & +50 12 20 & 3335   &  44.0  & 10.4	 & 34	   & 1.4   & (R)SA(r)a:?   & Sy1.9	    & 11,29\\ 
NGC~2782            & 09 14 05.1 & +40 06 49 & 2560   &  34.0  & 10.5	 & 47	   & 1.1   & SAB(rs)a      & Sy1,SB	    & 18\\ 
NGC~2824 (Mrk~394)  & 09 19 02.2 & +26 16 12 & 2760   &  37.0  & 9.9	 & 47	   & 2.7   & S0            & Sy? 	    & 30\\ 
SBS~0927+493        & 09 31 06.7 & +49 04 47 & 10167  & 135.6  & 11.2    & 52      &2.7$^c$&               & L$^g$          & 19\\	
UGC~5101 	    & 09 35 51.6 & +61 21 11 & 11809  & 157.5  & 12.0    & 46	   & 3.2   & S?            & Sy1.5,L,LIRG   & 27\\
NGC~2960 (Mrk~1419) & 09 40 36.4 & +03 34 37 & 4930   &  66.0  & 10.4	 & 40	   & 2.6   & Sa?           & L$^g$  	    & 31\\ 
NGC~2979            & 09 43 08.5 &--10 23 01 & 2720   &  36.0  & 10.0	 & 40	   & 2.1   & (R')SA(r)a?   & Sy2	    & 30\\ 
NGC~2989 	    & 09 45 25.8 &--18 22 36 &  4146  &  55.3  & 10.5    & 38	   & 1.6   & SAB(s)bc:     & HII	    & 1\\
NGC~3034 (M~82)     & 09 55 52.2 & +69 40 47 & 200    &  3.7   & 10.6	 & 65	   & 0.0   & I0            & SB,HII	    & 21,32\\ 
NGC~3079            & 10 01 57.8 & +55 40 47 & 1120   &  15.5  & 10.6	 & 42	   & 2.7   & SB(s)c        & Sy2,L	    & 33,34,35\\ 
IC~2560             & 10 16 18.7 &--33 33 50 & 2925   &  35.0  & 10.2	 & 47	   & 2.0   & (R')SB(r)bc   & Sy2,HII?	    & 24,36\\ 
Mrk~34              & 10 34 08.6 & +60 01 52 & 15140  &  205.0 & 11.3	 & 56	   & 3.0   & Spiral        & Sy2	    & 23\\ 
NGC~3359 	    & 10 46 36.8 & +63 13 25 &  1014  &  13.5  & 9.6  	 & 34	   & --0.2 & SB(rs)c       & HII	    & 1\\
NGC~3393 	    & 10 48 23.4 &--25 09 43 &  3750  &  50.0  & 10.4  	 & 46	   & 2.6   & (R')SB(s)ab   & Sy2	    & 5,27\\
NGC~3556            & 11 11 31.2 & +55 40 25 & 700    &  12.0  & 10.2	 & 38	   & 0.0   & SB(s)cd       & HII	    & 23\\ 
Arp~299 (Mrk~171)   & 11 28 32.2 & +58 33 44 & 3120   &  42.0  & 11.7	 & 61	   & 2.1   & IBm/SBm       &		    & 23\\ 
NGC~3735            & 11 35 57.3 & +70 32 09 & 2695   &  36.0  & 10.6	 & 38	   & 1.3   & SAc           & Sy2	    & 37\\ 
NGC~4051            & 12 03 09.6 & +44 31 53 & 730    &  10.0  & 9.6	 & 38	   & 0.3   & SAB(rs)bc     & Sy1.5          & 38\\ 
NGC~4151            & 12 10 32.6 & +39 24 21 & 1000   &  13.5  & 9.8     & 52      & --0.2 & (R')SAB(rs)ab:& Sy1.5          & 18\\ 
NGC~4258            & 12 18 57.5 & +47 18 14 & 450    &  7.2   & 9.9	 & 33	   & 1.9   & SAB(s)bc      & Sy1.9,L        & 21,39\\ 
NGC~4293 	    & 12 21 12.9 & +18 22 57 &  890   &  17    & 9.7  	 & 40	   & 0.7   & (R)SB(s)0/a   & L              & 5\\
NGC~4388            & 12 25 46.7 & +12 39 44 & 2520   &  34.0  & 10.7	 & 47	   & 1.1   & SA(s)b:       & Sy2            & 18\\ 
NGC~4527 	    & 12 34 08.5 & +02 39 14 & 1736   &  23.2  & 10.8  	 & 39	   & 0.6   & SAB(s)bc      & L,HII          & 1\\
ESO~269--G012       & 12 56 40.7 &--46 55 31 & 4950   &  66.0  & 10.5    & 41      & 3.0   & S0            & Sy2            & 30\\ 
NGC~4922            & 13 01 25.2 & +29 18 50 & 7080   &  95.0  & 11.2	 & 61	   & 2.3   & I0/p          & Sy2,L          & 18\\ 
NGC~4945            & 13 05 27.5 &--49 28 06 & 560    &  4.0   & 10.3	 & 45	   & 1.7   & SB(s)cd       & Sy2            & 40,41 \\ 
NGC~5194 (M~51a)     & 13 29 52.7 & +47 11 43 & 450    &  10.0  & 10.3	 & 33	   & --0.2 & SA(s)bc pec   & Sy2,HII        & 6,42\\ 
NGC~5256 (Mrk~266)  & 13 38 17.2 & +48 16 32 & 8365   & 112.0  & 11.5	 & 46	   & 1.5   & Pec           & Sy2,LIRG,SB    & 18\\ 
NGC~5347            & 13 53 17.8 & +33 29 27 & 2335   &  31.0  & 9.9	 & 44	   & 1.5   & (R')SB(rs)ab  & Sy2            & 24\\ 
NGC~5495 	    & 14 12 23.3 &--27 06 29 & 6589   &  87.8  & 10.8 	 & 39	   & 2.3   & (R')SA(rs)b   & Sy2,HII?       & 5\\
Circinus            & 14 13 09.3 &--65 20 21 & 450    &  4.0   & 10.1	 & 54	   & 1.3   & SA(s)b:       & Sy2            & 43,44\\ 
NGC~5506 (Mrk~1376) & 14 13 14.8 &--03 12 27 & 1850   &  25.0  & 10.3	 & 59	   & 1.7   & SA pec        & Sy1.9          & 11\\ 
NGC~5643            & 14 32 40.7 &--44 10 28 & 1200   &  16.0  & 10.3	 & 39	   & 1.4   & SAB(rs)c      & Sy2            & 30\\ 
NGC~5728            & 14 42 23.9 &--17 15 11 & 2795   &  37.0  & 10.6	 & 45	   & 1.9   & (R\_1)SAB(r)a & Sy2,HII        & 18\\ 
UGC~9618B 	    & 14 57 00.7 & +24 37 03 & 10094  &  134.6 & 11.7  	 & 39	   &3.2$^c$& (Sb)          & L,HII          & 19\\
NGC~5793            & 14 59 24.7 &--16 41 36 & 3490   &  47.0  & 10.6	 & 47	   & 2.0   & Sb:           & Sy2            & 45,46\\ 
NGC~6240            & 16 52 58.1 & +02 23 50 & 7340   &  98.0  & 11.8	 & 58	   & 1.6   & I0: pec       & Sy2,L          & 47,48,49,50\\
NGC~6264 	    & 16 57 16.1 & +27 50 59 & 10177  &  135.7 & \nodata & \nodata &3.1$^c$& Sb            & Sy2            & 19\\
NGC~6323            & 17 13 18.0 & +43 46 56 & 7790   &  104.0 & 10.3$^h$& 35$^h$  & 2.7   & Sab           & Sy2            & 18\\ 
NGC~6300            & 17 17 00.3 &--62 49 15 & 1110   &  15.0  & 10.2	 & 36	   & 0.5   & SB(rs)b       & Sy2            & 30\\ 
ESO~103--G035       & 18 38 20.3 &--65 25 42 & 3985   &  53.0  & 10.5	 & 121     & 2.6   & SA0           & Sy1,Sy2        & 24\\ 
IRAS~F19370--0131   & 19 39 38.9 &--01 24 33 & 6000   &  80.0  & 10.7	 & 59	   & 2.2   & Sb            & Sy2,HII        & 30\\ 
3C~403              & 19 52 15.8 & +02 30 24 & 17690  &  235.0 & 11.3$^i$&\nodata$^i$& 3.3   & S0            & NLRG         & 51\\ 
NGC~6926            & 20 31 38.7 &--80 49 58 & 5970   &  80.0  & 11.2	 & 39	   & 2.7   & SB(s)bc pec   & Sy2,HII        & 30\\ 
AM~2158--380b 	    & 22 01 17.1 &--37 46 24 & 9661   &  128.8 & 10.4$^h$& 49$^h$  & 2.7   & Sa            & Sy2,RG         & 5\\
TXS~2226--184       & 22 29 12.5 &--18 10 47 & 7495   &  100.0 & \nodata & \nodata & 3.8   & S?$^j$      & L                & 52\\ 
NGC~7479            & 23 04 56.7 & +12 19 22 & 2381   &  31.75 & 10.7    & 42      & 1.2   & SB(s)c        & Sy2,L          & 1\\ 
IC~1481             & 23 19 25.1 & +05 54 21 & 6120   &  82.0  & 10.6	 & 65	   & 2.5   & S?            & L              & 24\\
\enddata
\tablecomments{
Col. (1): Source. Cols. (2,3): RA and DEC (J2000). Col. (4): Systemic velocity c $z$ (km\,s$^{-1}$) taken
from NED.
Col. (5): Distance (Mpc), using $H_0$ = 75\,km\,s$^{-1}$\,Mpc$^{-1}$. For the 53 of the 78 masers that were included in \citet{hen05a},
we adapt the distances from their Table~4. For a few objects, we take distances 
from the references listed in Col.11. 
Col. (6, 7): Far-Infrared (FIR) luminosity log ($L_{\rm FIR}$/$L_{\odot}$) and dust temperature $T_{\rm dust}$ in K.
For the determination of $L_{\rm FIR}$ and $T_{\rm dust}$ (60/100~$\mu$m color temperatures), 
see \citet{wou86}. 
The IRAS fluxes were taken from \citet{ful89} and, 
for a few sources (NGC~262, NGC~598, IC~0184, NGC~4151, NGC~4258, IRAS~F19370--0131, NGC~6323, 3C~403, and AM~2158--380b), 
from NED. Col. (8): log ($L_{\rm H_2O}$/$L_{\odot}$), using $H_0$ = 75\,km\,s$^{-1}$\,Mpc$^{-1}$. 
Col. (9) Host galaxy morphology type taken from NED. Col. (10) 
Activity type taken from NED (Sy = Seyfert, L = LINER, LIRG = Luminous-Infrared Galaxy, 
SB = Starburst, NLRG = Narrow-Line Radio Galaxy, RG = Radio Galaxy).\\
Col. (11): References: 
[1] \citet{bra08};
[2] \citet{bec93};   
[3] \citet{arg94};   
[4] \citet{hen86};   
[5] \citet{kon06b};   
[6] \citet{ho87};    
[7] \citet{hen04};   
[8] \citet{fal00a};  
[9] \citet{pec03};   
[10] \citet{gre02};  
[11] \citet{bra94}; 
[12] \citet{cas08}; 
[13] \citet{chu77};  
[14] \citet{gre93};  
[15] \citet{huc78}; 
[16] \citet{bru06};
[17] \citet{huc88};
[18] \citet{bra04};  
[19] \citet{kon06a};  
[20] \citet{cla98};  
[21] \citet{cla84};  
[22] \citet{gal01};  
[23] \citet{hen05a};  
[24] \citet{bra96};  
[25] \citet{tar02a}; 
[26] \citet{tar02b}; 
[27] \citet{zha06};  
[28] \citet{bar05};  
[29] \citet{wil95};  
[30] \citet{gre03a}; 
[31] \citet{hen02};   
[32] \citet{bau96};  
[33] \citet{hen84};  
[34] \citet{has85};  
[35] \citet{tro98};  
[36] \citet{ish01};  
[37] \citet{gre97a}; 
[38] \citet{hag03b}; 
[39] \citet{her99};  
[40] \citet{dos79};  
[41] \citet{gre97b}; 
[42] \citet{hag01b}; 
[43] \citet{gar82};  
[44] \citet{gre03b}; 
[45] \citet{hag97};
[46] \citet{hag01a}; 
[47] \citet{hag02};  
[48] \citet{nak02};  
[49] \citet{bra03};  
[50] \citet{hag03a}; 
[51] \citet{tar03};  
[52] \citet{koe95}\\  
$^a$ While the detection of two more masers has been claimed  
by \citet{huc78,huc88}, 
we list only the three masers confirmed by \citet{bru06}.\\
$^b$ $L_{\rm FIR}$ and $T_{\rm dust}$
are lower limits as there are no 
IRAS flux measurements at 12 and 25$\mu$m.\\
$^c$ Estimated from Fig.~1 of \citet{kon06a} using 
$L_{\rm H_2O}/L_{\odot}$ = 0.023 $S_{\rm total}/({\rm Jy\,km\,s^{-1}})$ 
$\times$ $D^2/{\rm Mpc}$ 
with $S_{\rm total}$ = $\sum$ (1.06 $S_{\rm peak}$ $\times$ FWHM)
(summed over the different components).\\
$^d$ The maser luminosity refers to a brief flaring episode.\\
$^e$ We assume $\log L_{\rm H_2O}/L_{\odot}$ = 0.9 as total integrated
single-dish flux density when including NGC~2146 in Figs.~\ref{lf},\ref{dust},\ref{fir}.\\
$^f$ Using $H_0$ = 75\,km\,s$^{-1}$\,Mpc$^{-1}$, $\Omega_{\Lambda}$ = 0.73,
and $\Omega_{\rm M}$ = 0.27.\\
$^g$ Note that NED gives ``Sy3'' as classification, which corresponds to a LINER \citep{ver96}.\\
$^h$ There is no IRAS flux measurement at 12$\mu$m.\\
$^i$ There is no IRAS flux measurement at 100$\mu$m.\\
$^j$ Note that while NED lists ``E/S0'' as morphological classification of the host galaxy according to
\citet{koe95}, \citet{fal00b} favor a 
classification as a highly inclined spiral galaxy. We use the latter in our statistics in \textsection{5}.}
\label{maserlf}
\end{deluxetable}

\begin{deluxetable}{lcccc}
\tabletypesize{\scriptsize}
\tablecolumns{5}
\tablewidth{0pc}
\tablecaption{SDSS type-2 AGNs with IRAS flux measurements}
\tablehead{
\colhead{Source} & \colhead{$z$}
& \colhead{$D$} & \colhead{log $L_{\rm FIR}$} & \colhead{$T_{\rm dust}$}\\
&  & (Mpc) & ($L_{\odot}$) & (K)  \\
\colhead{(1)} & \colhead{(2)} & \colhead{(3)}  & \colhead{(4)} & \colhead{(5)}}
\startdata
022606.86--001656.0 & 0.407 & 2087.5 & 13.3 &  37\\
024919.01--000722.5 & 0.579 & 3193.5 & 13.7 &  37\\
031319.96+003715.6 & 0.395 & 2014.6 & 13.3 &  38\\ 
032240.60+001626.0 & 0.344 & 1711.3 & 13.1 &  35\\
075238.68+390304.9 & 0.654 & 3707.8 & 13.8 &  37\\
110709.36+511328.6 & 0.441 & 2297.4 & 13.5 &  32\\
123453.10+640510.2 & 0.594 & 3294.9 & 13.9 &  37\\
172603.09+602115.7 & 0.333 & 1647.4 & 13.1 &  35\\
\enddata
\tablecomments{
Col. (1): Source. 
Col. (2): Heliocentric redshift from \citet{zak03}
as measured from the [OII]\,$\lambda$3727 emission line. 
Col. (3): Distance (Mpc), using $H_0$ = 75\,km\,s$^{-1}$\,Mpc$^{-1}$,
$\Omega_{\Lambda}$ = 0.73, and $\Omega_{\rm M}$ = 0.27.
Col. (4, 5):  FIR luminosity log ($L_{\rm FIR}$/$L_{\odot}$) and dust temperature $T_{\rm dust}$ in K
(see also Table~\ref{maserlf} and text for details).
}
\label{zakfir}
\end{deluxetable}

\appendix

\section{ADDITIONAL OBSERVATIONS}
In addition to the SDSS type-2 AGNs, a sample of 76
radio galaxies at $z$ = 0.3--0.7 was observed
during the same observing runs at the GBT and the Effelsberg 100-m radio telescope
(Table~\ref{table2}).
An additional sample of high-redshift radio galaxies was observed at Arecibo Observatory
(see A1; Table~\ref{table2}).
The radio galaxies were selected from 
objects in NED classified as ``galaxy'' and ``radio source''
whose frequencies were accessible to the different telescope receivers
and whose declinations were in the observatory range.
Also, a few sources classified as galaxy or QSO were observed (26 in total; Table~\ref{table3}).
\subsection{Arecibo}
A total of 71 objects with $z$ = 1.1--3.6 were observed during observing runs in October 2003,
June and September 2004, and March 2005
in dual polarization, single beam, position-switching mode
(see Table~\ref{table2}). Each observation consisted of typically 5
on/off scans of 10 minutes each, resulting in a total duration of 50 minutes per source.
(However, some objects only have 2 or 3 individual scans.)
Total spectrometer bandwidths of either 25 MHz or 50 MHz
were divided into 1024 channels. Antenna gain, as determined from previous measurements by Arecibo staff, ranged
from about 3 K/Jy to 7 K/Jy, depending on frequency, and flux calibration was done using these values obtained from
a lookup table. We estimate that calibrations are accurate to ~20\%. 
Pointing was checked approximately every two
hours using extragalactic radio sources.

\begin{deluxetable}{lccccccccc}
\tabletypesize{\scriptsize}
\tablecolumns{10}
\tablewidth{0pc}
\tablecaption{Details of Observations: Radio Galaxies}
\tablehead{
\colhead{Source} & \colhead{RA} & \colhead{DEC}& \colhead{$z$} & \colhead{$\nu$} & \colhead{rms} & 
\colhead{$v$ range} & \colhead{Channel} & \colhead{Telescope} & \colhead{Epoch}\\
& & & & & & & width\\
& \multicolumn{2}{c}{(J2000)} & & (MHz) & (mJy) & (km\,s$^{-1}$) & (km\,s$^{-1}$) & & \\
\colhead{(1)} & \colhead{(2)} & \colhead{(3)}  & \colhead{(4)} & \colhead{(5)}
& \colhead{(6)} & \colhead{(7)} & \colhead{(8)} & \colhead{(9)} & \colhead{(10)}}
\startdata
RC~0008+172             & 00 11 07.5 & +17 29 48 & 1.390 &  9303.35 & 2.8  & --820,805	 & 1.5 & Arecibo    & 1003\\
3C~005                  & 00 13 10.5 & +00 51 32 & 0.606 & 13858.80 & 1.4  & --2160,2160 & 0.5 & GBT        & 0605\\
PMN~J0018+0940          & 00 18 55.2 & +09 40 07 & 1.586 &  8598.22 & 2.2  & --870,870	 & 1.7 & Arecibo    & 1003\\
B2~0026+34              & 00 29 14.2 & +34 56 32 & 0.517 & 14657.30 & 2.1  & --2045,2045 & 0.5 & GBT	    & 0605\\ 
4C~+45.02     		& 00 30 52.1 & +45 21 48 & 0.365 & 16289.00 & 12.5 & --1670,1720 & 1.4 & Effelsberg & 1205\\
3C~016	     		& 00 37 44.6 & +13 19 55 & 0.405 & 15826.00 & 11.7 & --1720,1780 & 1.5 & Effelsberg & 1205\\
PKS~0037--009           & 00 40 20.3 &--00 40 33 & 0.568 & 14180.50 & 2.0  & --2110,2110 & 0.5 & GBT        & 0605\\
3C~19                   & 00 40 55.0 & +33 10 08 & 0.482 & 15003.40 & 1.9  & --1995,1995 & 0.5 & GBT        & 0605\\
MRC~0044+107            & 00 46 41.4 & +11 02 53 & 1.813 &  7904.37 & 0.7  & --950,950	 & 1.8 & Arecibo    & 0904\\
MG3~J005335+2045        & 00 53 37.9 & +20 46 03 & 1.297 &  9680.02 & 3.3  & --770,770	 & 1.5 & Arecibo    & 1003\\
4C~+09.03               & 00 57 29.7 & +09 17 54 & 1.301 &  9663.19 & 2.9  & --775,775	 & 1.5 & Arecibo    & 1003\\
PKS~0101+023  		& 01 04 23.9 & +02 39 43 & 0.390 & 15996.00 & 10.4 & --1700,1760 & 1.5 & Effelsberg & 1205\\
3C~34                   & 01 10 18.6 & +31 47 20 & 0.690 & 13156.80 & 2.1  & --2275,2275 & 0.6 & GBT        & 0605\\
4C~08.06                & 01 19 01.3 & +08 29 55 & 0.594 & 13952.60 & 2.0  & --2145,2145 & 0.5 & GBT        & 0605\\
NVSS~J012142+132058     & 01 21 42.7 & +13 20 58 & 3.516 &  4923.61 & 0.7  & --1520,1520 & 2.9 & Arecibo    & 1003\\
MG1~J012232+1923        & 01 22 29.9 & +19 23 39 & 1.595 &  8568.40 & 1.6  & --875,875	 & 1.7 & Arecibo    & 1003\\
3C~44                   & 01 31 21.7 & +06 23 41 & 0.660 & 13394.60 & 2.1  & --2235,2235 & 0.5 & GBT        & 0605\\
3C~45                   & 01 35 15.0 & +08 11 08 & 0.499 & 14833.30 & 1.5  & --2020,2020 & 0.5 & GBT        & 0605\\
87GB~013419.5+250603    & 01 37 06.7 & +25 21 19 & 2.897 &  5705.67 & 1.0  & --1315,1315 & 2.5 & Arecibo    & 1003\\
4C~--01.09              & 01 43 17.3 &--01 18 58 & 0.520 & 14628.30 & 1.5  & --2045,2045 & 0.5 & GBT        & 0605\\
3C~55                   & 01 57 10.5 & +28 51 38 & 0.735 & 12831.50 & 1.3  & --2335,2335 & 0.6 & GBT        & 0605\\
NVSS~J020510+224250     & 02 05 10.7 & +22 42 51 & 3.056 &  4934.42 & 0.9  & --1520,1520 & 2.9 & Arecibo    & 1003\\
87GB~021027.8+325342    & 02 13 27.1 & +33 08 03 & 1.456 &  9053.34 & 2.2  & --825,825	 & 1.6 & Arecibo    & 1003\\
NVSS~J023111+360027     & 02 31 11.7 & +36 00 27 & 3.079 &  5451.09 & 1.0  & --1375,1375 & 2.6 & Arecibo    & 1003\\
3C~068.2                & 02 34 23.8 & +31 34 17 & 1.575 &  8634.95 & 2.4  & --870,870	 & 1.7 & Arecibo    & 1003\\
MG3~J025952+3237        & 02 59 55.3 & +32 37 42 & 1.657 &  8368.46 & 1.9  & --890,880	 & 1.7 & Arecibo    & 1003\\
074324.07+233626.0  	& 07 43 24.1 & +23 36 26 & 0.417 & 15695.00 & 11.3 & --1720,1790 & 1.5 & Effelsberg & 1205\\
074556.44+191306.2  	& 07 45 56.4 & +19 13 06 & 0.388 & 16025.00 & 9.9  & --1700,1750 & 1.5 & Effelsberg & 1205\\
074654.59+193413.7     	& 07 46 54.6 & +19 34 14 & 0.385 & 16054.00 & 7.8  & --1720,1750 & 1.5 & Effelsberg & 1205\\
MRC~0748+134            & 07 51 01.2 & +13 19 20 & 2.419 &  6503.36 & 1.0  & --575,575	 & 1.1 & Arecibo    & 0305\\
B2~0817+27   		& 08 20 23.9 & +27 43 07 & 0.347 & 16513.00 & 12.7 & --1640,1700 & 1.4 & Effelsberg & 1205\\
B2~0820+36B             & 08 23 48.1 & +36 32 46 & 1.860 &  7774.47 & 1.3  & --480,480	 & 0.9 & Arecibo    & 0305\\
4C~+34.28     		& 08 25 12.2 & +34 19 06 & 0.406 & 15814.00 & 10.5 & --1720,1780 & 1.5 & Effelsberg & 1205\\
MG2~J082558+2643        & 08 25 57.0 & +26 43 58 & 1.609 &  8522.42 & 0.8  & --440,440	 & 0.9 & Arecibo    & 0305\\
B3~0825+375  		& 08 28 19.7 & +37 21 53 & 0.389 & 16013.00 & 9.9  & --1690,1750 & 1.5 & Effelsberg & 1205\\
B2~0825+34              & 08 28 26.7 & +34 42 50 & 1.460 &  9038.62 & 1.5  & --415,415	 & 0.9 & Arecibo    & 0305\\
87GB~082540.4+253730    & 08 28 39.7 & +25 27 30 & 2.218 &  6909.57 & 1.0  & --540,540	 & 1.1 & Arecibo    & 0305\\
MG2~J082900+2453        & 08 28 59.5 & +24 54 00 & 2.224 &  6896.71 & 1.1  & --170,540	 & 1.1 & Arecibo    & 0305\\
TXS~0828+193            & 08 30 53.4 & +19 13 16 & 2.572 &  6224.80 & 1.5  & --600,600	 & 1.2 & Arecibo    & 0305\\
4C~+14.27     		& 08 35 03.6 & +14 11 49 & 0.392 & 15973.00 & 18.1 & --1700,1760 & 1.5 & Effelsberg & 1205\\
083825.01+371036.6  	& 08 38 25.0 & +37 10 37 & 0.396 & 15928.00 & 9.4  & --1710,1770 & 1.5 & Effelsberg & 1205\\
084137.59+311247.6  	& 08 41 37.6 & +31 12 48 & 0.335 & 16660.00 & 16.7 & --1680,1620 & 1.4 & Effelsberg & 1205\\
4C~+29.31     		& 08 43 09.9 & +29 44 05 & 0.398 & 15905.00 & 10.1 & --1700,1760 & 1.5 & Effelsberg & 1205\\
B2~0856+33              & 08 59 32.0 & +33 04 12 & 0.497 & 14843.20 & 2.4  & --2015,2015 & 0.5 & GBT        & 0605\\
091050.78+012212.8  	& 09 10 50.8 & +01 22 13 & 0.334 & 16670.00 & 11.8 & --1680,1620 & 1.4 & Effelsberg & 1205\\
CL~09104+4109           & 09 13 36.6 & +40 56 35 & 0.442 & 15419.50 & 1.4  & --1940,1940 & 0.5 & GBT	    & 0605\\ 
093357.76+485315.1  	& 09 33 57.8 & +48 53 15 & 0.385 & 16059.00 & 13.3 & --1690,1750 & 1.5 & Effelsberg & 1205\\
3C~222                  & 09 36 32.0 & +04 22 10 & 1.339 &  9506.20 & 1.6  & --390,390	 & 0.8 & Arecibo    & 0305\\
B3~0934+387  		& 09 37 53.4 & +38 33 37 & 0.358 & 16370.00 & 9.7  & --1650,1700 & 1.4 & Effelsberg & 1205\\
4C~--01.19     		& 09 41 22.6 &--01 43 01 & 0.382 & 16089.00 & 11.9 & --1690,1730 & 1.5 & Effelsberg & 1205\\
3C~230                  & 09 51 58.8 &--00 01 27 & 1.487 &  8940.49 & 1.8  & --420,420	 & 0.8 & Arecibo    & 0305\\
2MASX~J09542562--0055429& 09 54 25.6 &--00 55 43 & 0.358 & 16368.00 & 10.9 & --1680,1720 & 1.4 & Effelsberg & 1205\\
095718.57+123955.8  	& 09 57 18.6 & +12 39 56 & 0.395 & 15936.00 & 9.4  & --1700,1760 & 1.5 & Effelsberg & 1205\\
100531.61+023508.2  	& 10 05 31.6 & +02 35 08 & 0.370 & 16235.00 & 12.4 & --1680,1730 & 1.4 & Effelsberg & 1205\\
3C~238                  & 10 11 00.4 & +06 24 40 & 1.405 &  9245.00 & 1.5  & --405,405	 & 0.8 & Arecibo    & 0305\\
4C~+05.41               & 10 19 33.4 & +05 34 34 & 2.765 &  5905.71 & 1.0  & --635,635	 & 1.2 & Arecibo    & 0305\\
B2 1016+36	        & 10 19 53.4 & +36 22 46 & 1.892 &  7688.45 & 1.3  & --487,487	 & 0.9 & Arecibo    & 0305\\
3C~241                  & 10 21 54.5 & +21 59 30 & 1.617 &  8496.37 & 1.0  & --350,350	 & 0.9 & Arecibo    & 0305\\
4C~+39.32     		& 10 28 44.3 & +38 44 37 & 0.361 & 16337.00 & 10.4 & --1670,1720 & 1.4 & Effelsberg & 1205\\
TXS~1030+074            & 10 33 34.0 & +07 11 26 & 1.540 &  8753.94 & 1.1  & --430,430	 & 0.8 & Arecibo    & 0305\\
4C~+34.32               & 10 34 35.1 & +33 49 21 & 1.832 &  7851.34 & 2.3  & --305,475	 & 0.9 & Arecibo    & 0305\\
B3~1046+465  		& 10 49 55.4 & +46 19 12 & 0.351 & 16458.00 & 12.1 & --1660,1710 & 1.4 & Effelsberg & 1205\\
B3~1051+454  		& 10 54 00.1 & +45 08 05 & 0.384 & 16065.00 & 13.0 & --1700,1750 & 1.5 & Effelsberg & 1205\\
4C~+08.34               & 11 07 09.4 & +08 41 33 & 1.384 &  9326.76 & 1.3  & --400,400	 & 0.8 & Arecibo    & 0305\\
4C~+34.34               & 11 16 30.4 & +34 42 24 & 2.400 &  6539.70 & 1.2  & --570,570	 & 1.1 & Arecibo    & 0305\\
3C~256                  & 11 20 43.0 & +23 27 55 & 1.819 &  7887.54 & 1.3  & --470,470	 & 0.9 & Arecibo    & 0305\\
7C~1121+3157            & 11 23 55.8 & +31 41 29 & 3.217 &  5272.21 & 1.1  & --235,235	 & 1.4 & Arecibo    & 0305\\
MRC~1139+139            & 11 42 23.7 & +13 38 01 & 1.279 &  9756.47 & 1.8  & --385,385	 & 0.8 & Arecibo    & 0305\\
4C~+35.26               & 11 43 51.1 & +35 08 23 & 1.781 &  7995.32 & 1.0  & --470,470	 & 0.9 & Arecibo    & 0305\\
4C~+26.38               & 12 32 23.6 & +26 04 07 & 2.608 &  6162.69 & 1.7  & --610,610	 & 1.2 & Arecibo    & 0305\\
4C~+34.37               & 12 32 41.4 & +34 42 50 & 1.533 &  8778.12 & 1.3  & --430,430	 & 0.8 & Arecibo    & 0305\\
4C~+03.24               & 12 45 38.3 & +03 23 21 & 3.570 &  4865.53 & 1.4  & --770,770	 & 1.5 & Arecibo    & 0305\\
MRC~1248+113            & 12 51 00.0 & +11 04 22 & 2.332 &  6693.25 & 0.7  & --1075,1075 & 2.2 & Arecibo    & 0604\\
4C~+09.45               & 13 05 36.0 & +08 55 14 & 1.409 &  9229.97 & 1.0  & --810,810	 & 1.6 & Arecibo    & 0604\\
PMN~J1401+0921          & 14 01 18.5 & +09 21 21 & 2.093 &  7188.81 & 0.7  & --1040,1040 & 2.0 & Arecibo    & 0604\\
B3~1402+415  		& 14 04 16.4 & +41 17 49 & 0.360 & 16346.00 & 10.6 & --1670,1720 & 1.4 & Effelsberg & 1205\\
3C~294                  & 14 06 44.0 & +34 11 25 & 1.779 &  8001.08 & 1.1  & --790,940	 & 1.8 & Arecibo    & 0604\\
3C~299	     		& 14 21 05.6 & +41 44 48 & 0.367 & 16266.00 & 12.2 & --1670,1720 & 1.4 & Effelsberg & 1205\\
MRC~1436+157            & 14 39 05.0 & +15 31 19 & 2.538 &  6284.00 & 0.6  & --1190,1190 & 2.2 & Arecibo    & 0604\\
B2~1459+37   		& 15 01 19.1 & +37 27 42 & 0.340 & 16598.00 & 15.4 & --1640,1700 & 1.4 & Effelsberg & 1205\\
4C~+03.31               & 15 05 06.5 & +03 47 11 & 1.652 &  8384.24 & 1.0  & --895,895	 & 1.7 & Arecibo    & 0604\\
150609.62+505345.4  	& 15 06 09.6 & +50 53 45 & 0.411 & 15762.00 & 13.2 & --1740,1780 & 1.5 & Effelsberg & 1205\\
3C~313                  & 15 11 00.0 & +07 51 50 & 0.461 & 15219.10 & 1.9  & --1970,1970 & 0.5 & GBT        & 0605\\
4C~+24.33               & 15 18 39.9 & +24 27 02 & 1.847 &  7809.97 & 0.8  & --960,960	 & 1.8 & Arecibo    & 0604\\
3C~318                  & 15 20 05.4 & +20 16 06 & 1.574 &  8638.33 & 1.5  & --870,870	 & 1.7 & Arecibo    & 0604\\
4C~+04.51               & 15 21 14.4 & +04 30 22 & 1.296 &  9684.23 & 1.1  & --775,775	 & 1.5 & Arecibo    & 0604\\   
4C~+04.52               & 15 22 32.8 & +04 00 30 & 0.534 & 14494.80 & 1.9  & --2065,2065 & 0.5 & GBT        & 0605\\
3C~323                  & 15 41 45.5 & +60 15 35 & 0.679 & 13243.10 & 2.2  & --2260,2260 & 0.6 & GBT        & 0605\\
3C~324                  & 15 49 48.9 & +21 25 38 & 1.206 & 10079.36 & 1.6  & --745,745	 & 1.4 & Arecibo    & 0604\\
3C~326.1                & 15 56 10.1 & +20 04 20 & 1.825 &  7870.79 & 0.6  & --955,955	 & 1.8 & Arecibo    & 0604\\
3C~327.1                & 16 04 45.3 & +01 17 51 & 0.462 & 15208.70 & 2.6  & --1970,1970 & 0.5 & GBT        & 0605\\
3C~330                  & 16 09 36.6 & +65 56 44 & 0.550 & 14345.20 & 3.2  & --2090,2090 & 0.5 & GBT        & 0105\\
                        &            &           &       & 14345.20 & 2.2  & --2090,2090 & 0.5 & GBT        & 0605\\ 
4C~+06.56     		& 16 21 32.5 & +06 07 19 & 0.343 & 16556.00 & 14.9 & --1600,1700 & 1.4 & Effelsberg & 1205\\
3C~337                  & 16 28 52.8 & +44 19 05 & 0.635 & 13599.40 & 2.1  & --2200,2200 & 0.5 & GBT        & 0605\\
4C~+12.60               & 16 40 47.9 & +12 20 02 & 1.152 & 10332.29 & 1.7  & --725,725	 & 1.4 & Arecibo    & 0604\\
3C~344                  & 16 43 05.6 & +37 29 13 & 0.520 & 14641.60 & 1.1  & --2045,2045 & 0.5 & GBT        & 0605\\
PMN~J1650+0955          & 16 50 05.1 & +09 55 09 & 2.509 &  6336.56 & 0.8  & --1180,1180 & 2.2 & Arecibo    & 0604\\
3C~349	     		& 16 59 29.5 & +47 02 44 & 0.205 & 18452.00 & 16.6 & --1520,1460 & 1.3 & Effelsberg & 1205\\
B3~1701+423             & 17 02 55.9 & +42 17 49 & 0.476 & 15064.40 & 1.8  & --1990,1990 & 0.5 & GBT	    & 0605\\ 
PKS~1706+006            & 17 08 44.6 & +00 35 10 & 0.449 & 15345.10 & 1.6  & --1950,1950 & 0.5 & GBT        & 0605\\
4C~+10.48               & 17 10 06.5 & +10 31 06 & 2.349 &  6639.29 & 0.8  & --1130,1130 & 2.2 & Arecibo    & 0604\\
PKS1716+006             & 17 19 22.9 & +00 37 09 & 0.704 & 13048.80 & 2.3  & --2295,2295 & 0.6 & GBT        & 0605\\
4C~+16.48               & 17 27 35.5 & +16 44 25 & 1.508 &  8865.63 & 1.0  & --845,845	 & 1.6 & Arecibo    & 0604\\
4C~+56.25               & 17 30 38.5 & +56 42 46 & 0.640 & 13558.00 & 1.4  & --2210,2210 & 0.5 & GBT        & 0605\\
4C~+73.16     		& 17 32 03.7 & +73 40 04 & 0.226 & 18136.00 & 16.6 & --1550,1490 & 1.3 & Effelsberg & 1205\\
4C~+71.17     		& 17 45 43.5 & +71 15 49 & 0.216 & 18285.00 & 22.1 & --1540,1480 & 1.3 & Effelsberg & 1205\\
3C~362                  & 17 47 07.0 & +18 21 10 & 2.281 &  6776.89 & 1.0  &   0,885	 & 2.1 & Arecibo    & 0604\\
NEP~3450     		& 17 55 57.9 & +63 14 09 & 0.386 & 16019.00 & 10.9 & --1690,1760 & 1.5 & Effelsberg & 1205\\
4C~+13.66               & 18 01 38.9 & +13 51 24 & 1.450 &  9075.51 & 1.2  & --825,825	 & 1.6 & Arecibo    & 0604\\
3C~368                  & 18 05 06.3 & +11 01 33 & 1.131 & 10434.11 & 1.6  & --720,720	 & 1.4 & Arecibo    & 0604\\
4C~+70.20     		& 18 06 47.0 & +70 44 45 & 0.204 & 18468.00 & 16.6 & --1530,1460 & 1.3 & Effelsberg & 1205\\
87GB~180705.8+683108    & 18 06 50.1 & +68 31 43 & 0.580 & 14072.80 & 2.0  & --2130,2130 & 0.5 & GBT	    & 0605\\ 
4C~+68.20     		& 18 15 24.8 & +68 06 32 & 0.230 & 18077.00 & 18.4 & --1540,1490 & 1.3 & Effelsberg & 1205\\
87GB~182301.2+660210	& 18 23 02.8 & +66 03 25 & 0.370 & 16230.00 & 12.9 & --1660,1740 & 1.4 & Effelsberg & 1205\\
3C~379.1      		& 18 24 33.0 & +74 20 59 & 0.256 & 17703.00 & 26.1 & --1580,1530 & 1.3 & Effelsberg & 1205\\
87GB~182631.5+651051    & 18 26 41.7 & +65 12 43 & 0.646 & 13508.60 & 2.1  & --2215,2215 & 0.5 & GBT	    & 0605\\ 
87GB~183438.3+620153    & 18 35 10.9 & +62 04 08 & 0.519 & 14634.10 & 1.9  & --2045,2045 & 0.5 & GBT	    & 0605\\ 
87GB~200332.8+661716    & 20 03 54.5 & +66 25 56 & 0.456 & 15271.40 & 2.0  & --1960,1960 & 0.5 & GBT	    & 0605\\ 
PKS2008--068            & 20 11 14.2 &--06 44 04 & 0.547 & 14386.50 & 2.1  & --2080,2080 & 0.5 & GBT        & 0605\\
3C~410	     		& 20 20 06.5 & +29 42 14 & 0.249 & 17809.00 & 23.1 & --1580,1540 & 1.3 & Effelsberg & 1205\\
4C~+02.51               & 20 36 34.8 & +02 56 54 & 2.130 &  7103.83 & 1.1  & --590,250	 & 2.0 & Arecibo    & 0904\\
PMN~J2037-0010	        & 20 37 13.4 &--00 10 59 & 1.512 &  8851.51 & 7.2  & --844,844	 & 1.6 & Arecibo    & 0904\\
B2~2050+36   		& 20 52 52.1 & +36 35 35 & 0.354 & 16422.00 & 13.3 & --1680,1710 & 1.4 & Effelsberg & 1205\\
2MASX~J20580405--0018416& 20 58 04.1 &--00 18 42 & 0.336 & 16643.00 & 16.2 & --1630,1700 & 1.4 & Effelsberg & 1205\\
PMN~J2058+0542          & 20 58 28.8 & +05 42 51 & 1.381 &  9338.51 & 4.3  & --800,800	 & 1.5 & Arecibo    & 0904\\
3C~427.1                & 21 04 06.4 & +76 33 12 & 0.572 & 14158.10 & 1.9  & --2115,2115 & 0.5 & GBT        & 0605\\
4C~+23.56               & 21 07 14.8 & +23 31 45 & 2.483 &  6383.86 & 1.3  & --1080,320  & 2.2 & Arecibo    & 0904\\
4C~+03.49               & 21 09 21.8 & +03 26 52 & 1.634 &  8441.53 & 2.9  & --890,890	 & 1.7 & Arecibo    & 0904\\
87GB~211905.7+182558    & 21 21 25.5 & +18 39 09 & 1.861 &  7771.75 & 1.0  & --930,930	 & 1.9 & Arecibo    & 0904\\
3C~434	     		& 21 23 16.3 & +15 48 06 & 0.322 & 16819.00 & 26.5 & --1670,1620 & 1.4 & Effelsberg & 1205\\
3C~435                  & 21 29 06.1 & +07 33 06 & 0.471 & 15115.60 & 1.4  & --1980,1980 & 0.5 & GBT        & 0605\\
4C~+19.71               & 21 44 07.5 & +19 29 15 & 3.594 &  4840.00 & 1.0  & --1550,1550 & 1.0 & Arecibo    & 1003\\
3C~436	     		& 21 44 11.7 & +28 10 18 & 0.215 & 18308.00 & 24.3 & --1540,1480 & 1.3 & Effelsberg & 1205\\
3C~437                  & 21 47 25.1 & +15 20 37 & 1.480 &  8965.73 & 2.9  & --835,835	 & 1.6 & Arecibo    & 1003\\
4C~+26.59               & 21 58 25.4 & +26 52 37 & 0.713 & 12980.20 & 1.6  & --2310,2310 & 0.6 & GBT        & 0605\\
MRC~2202+128            & 22 05 14.3 & +13 05 32 & 2.706 &  5999.73 & 1.1  & --1250,1250 & 2.4 & Arecibo    & 1003\\
TXS~2226+162            & 22 28 43.9 & +16 33 16 & 1.519 &  8826.91 & 2.9  & --850,850	 & 1.6 & Arecibo    & 1003\\
4C~+11.68     		& 22 29 57.5 & +11 27 36 & 0.238 & 17960.00 & 18.3 & --1570,1510 & 1.3 & Effelsberg & 1205\\
4C~--04.85              & 22 39 32.8 &--04 29 34 & 0.552 & 14340.20 & 1.4  & --2090,2090 & 0.5 & GBT        & 0605\\
TXS~2252+186            & 22 54 53.7 & +18 57 02 & 2.153 &  7052.01 & 0.9  & --510,1020  & 2.0 & Arecibo    & 0904\\
PKS~2305+033            & 23 08 25.1 & +03 37 04 & 2.457 &  6431.87 & 0.9  & --560,1170  & 2.2 & Arecibo    & 0904\\
PMN~J2320+0500          & 23 20 44.7 & +04 59 30 & 0.635 & 13599.40 & 2.2  & --2204,2204 & 0.5 & GBT	    & 0605\\ 
87GB~231912.7+222125    & 23 21 42.2 & +22 37 52 & 2.553 &  6258.09 & 0.8  & --1200,1200 & 2.3 & Arecibo    & 0904\\
MRC~2332+154            & 23 34 58.4 & +15 45 50 & 2.480 &  6389.36 & 0.7  & --1120,0	 & 2.2 & Arecibo    & 0904\\
3C~467                  & 23 48 29.6 & +18 44 05 & 0.632 & 13624.40 & 2.1  & --2200,2200 & 0.5 & GBT        & 0605\\
4C~+10.74               & 23 51 26.9 & +10 34 54 & 1.334 &  9526.56 & 2.5  & --785,785	 & 1.6 & Arecibo    & 1003\\
4C~+28.58               & 23 51 59.2 & +29 10 29 & 2.891 &  5714.47 & 1.0  & --1310,1310 & 2.5 & Arecibo    & 1003\\
\enddata
\tablecomments{
Col. (1): Source. 
Col. (2,3): Right ascension  and declination (J2000) taken from NED.
Col. (4): Heliocentric redshift $z$ taken from NED. 
Col. (5): Observed Frequency $\nu$ in MHz. 
Col. (6): rms flux density in mJy.
Col. (7): Velocity range covered by observations in km\,s$^{-1}$.
Col. (8): Channel width in km\,s$^{-1}$.
Col. (9): Telescope at which the source was observed.
Col. (10): Date of observations (mmyy).
}
\label{table2}
\end{deluxetable}

\begin{deluxetable}{lcccccccccc}
\rotate
\tabletypesize{\scriptsize}
\tablecolumns{11}
\tablewidth{0pc}
\tablecaption{Details of Observations: Miscellaneous Sources}
\tablehead{
\colhead{Source} & \colhead{Classification} & \colhead{RA} & \colhead{DEC}& \colhead{$z$} & \colhead{$\nu$} & \colhead{rms} & 
\colhead{$v$ range} & \colhead{Channel} & \colhead{Telescope} & \colhead{Epoch}\\
& & & & & & & & width\\
& & \multicolumn{2}{c}{(J2000)} & & (MHz) & (mJy) & (km\,s$^{-1}$) & (km\,s$^{-1}$) & & \\
\colhead{(1)} & \colhead{(2)} & \colhead{(3)}  & \colhead{(4)} & \colhead{(5)}
& \colhead{(6)} & \colhead{(7)} & \colhead{(8)} & \colhead{(9)} & \colhead{(10)} & \colhead{(11)}}
\startdata
081404.54+060238.3  	& QSO & 08 14 04.5 & +06 02 38 & 0.561 & 14235.00 & 2.3  & --2105,2105 & 0.5 & GBT	  & 0605\\
082110.76+031758.4  	& QSO & 08 21 10.8 & +03 17 58 & 0.451 & 15324.00 & 3.3  & --1955,1955 & 0.5 & GBT	  & 0605\\
083134.21+290239.5  	& G   & 08 31 34.2 & +29 02 40 & 0.568 & 14180.50 & 1.5  & --2110,2110 & 0.5 & GBT	  & 0605\\
091345.49+405628.2  	& G   & 09 13 45.5 & +40 56 28 & 0.442 & 15430.30 & 2.2  & --1940,1940 & 0.5 & GBT	  & 0605\\
092344.07+081051.2  	& QSO & 09 23 44.1 & +08 10 51 & 0.415 & 15703.00 & 12.7 & --1730,1780 & 1.5 & Effelsberg & 1205\\
093737.11+370535.4  	& G   & 09 37 37.1 & +37 05 35 & 0.449 & 15334.60 & 2.2  & --1955,1955 & 0.5 & GBT	  & 0605\\ 
095005.83+481338.6  	& G   & 09 50 05.8 & +48 13 39 & 0.375 & 16171.00 & 11.9 & --1680,1730 & 1.4 & Effelsberg & 1205\\ 
095019.90+051140.9  	& G   & 09 50 19.9 & +05 11 41 & 0.523 & 14590.00 & 2.0  & --2055,2055 & 0.5 & GBT	  & 0605\\
100258.68+050812.0  	& G   & 10 02 58.7 & +05 08 12 & 0.512 & 14705.70 & 2.0  & --2035,2035 & 0.5 & GBT	  & 0605\\ 
101401.59+431543.4  	& G   & 10 14 01.6 & +43 15 43 & 0.511 & 14715.50 & 2.0  & --2035,2035 & 0.5 & GBT	  & 0605\\
102853.93+514323.1  	& G   & 10 28 53.9 & +51 43 23 & 0.417 & 15681.00 & 11.3 & --1710,1800 & 1.5 & Effelsberg & 1205\\ 
103822.07+523115.9  	& QSO & 10 38 22.1 & +52 31 16 & 0.598 & 13914.30 & 2.1  & --2150,2150 & 0.5 & GBT	  & 0605\\
112907.10+575605.4  	& G   & 11 29 07.1 & +57 56 05 & 0.312 & 16935.00 & 10.0 & --1660,1600 & 1.4 & Effelsberg & 1205\\ 
113710.78+573158.8  	& G   & 11 37 10.8 & +57 31 59 & 0.395 & 15939.00 & 7.5  & --1690,1750 & 1.5 & Effelsberg & 1205\\ 
2MASS~J12324114+1112587 & G   & 12 32 41.1 & +11 12 59 & 0.249 & 17802.31 & 1.3  & --500,1100  & 1.3 & Effelsberg & 1105\\
134021.43+564319.3  	& QSO & 13 40 21.4 & +56 43 19 & 0.572 & 14144.50 & 1.9  & --2115,2115 & 0.5 & GBT	  & 0605\\
141318.96+415658.8  	& G   & 14 13 19.0 & +41 56 59 & 0.533 & 14504.30 & 2.1  & --2065,2065 & 0.5 & GBT	  & 0605\\
141918.90+510240.2  	& G   & 14 19 18.9 & +51 02 40 & 0.324 & 16794.00 & 26.0 & --1680,1620 & 1.4 & Effelsberg & 1205\\ 
142939.80+395935.3  	& G   & 14 29 39.8 & +39 59 35 & 0.531 & 14523.20 & 2.0  & --2060,2060 & 0.5 & GBT	  & 0605\\
143017.34+521735.1  	& QSO & 14 30 17.4 & +52 17 35 & 0.367 & 16254.00 & 12.2 & --1680,1720 & 1.4 & Effelsberg & 1205\\ 
153944.13+343503.9  	& QSO & 15 39 44.1 & +34 35 04 & 0.551 & 14336.00 & 2.0  & --2090,2090 & 0.5 & GBT	  & 0605\\
155143.91+434758.1  	& G   & 15 51 43.9 & +43 47 58 & 0.618 & 13733.90 & 1.7  & --2180,2180 & 0.5 & GBT	  & 0605\\
161401.64+380807.5  	& G   & 16 14 01.6 & +38 08 08 & 0.350 & 16470.00 & 13.4 & --1630,1700 & 1.4 & Effelsberg & 1205\\ 
162459.70+332253.4  	& QSO & 16 24 59.7 & +33 22 53 & 0.504 & 14784.00 & 1.4  & --2025,2025 & 0.5 & GBT	  & 0605\\ 
163653.37+245746.4  	& QSO & 16 36 53.4 & +24 57 46 & 0.584 & 14028.50 & 1.5  & --2135,2135 & 0.5 & GBT	  & 0605\\ 
164424.58+344636.9  	& QSO & 16 44 24.6 & +34 46 37 & 0.587 & 14010.80 & 1.6  & --2135,2135 & 0.5 & GBT	  & 0605\\ 
\enddata
\tablecomments{
Col. (1): Source. 
Col. (2): Classification as Galaxy (G) or QSO, taken from NED.
Col. (3,4): Right ascension  and declination (J2000) taken from NED.
Col. (5): Heliocentric redshift $z$ taken from NED.
Col. (6): Observed Frequency $\nu$ in MHz. 
Col. (7): rms flux density in mJy.
Col. (8): Velocity range covered by observations in km\,s$^{-1}$.
Col. (9): Channel width in km\,s$^{-1}$.
Col. (10): Telescope at which the source was observed.
Col. (11): Date of observations (mmyy).}
\label{table3}
\end{deluxetable}


\begin{thebibliography}{}
\bibitem[Antonucci(1993)]{ant93} Antonucci, R. R. J. 1993, ARA\&A, 31, 473
\bibitem[Argon et al.(1994)]{arg94}  
Argon, A. L., Greenhill, L. J., Moran, J. M.,
Reid, M. J., Menten, K. M., Henkel, C., Inoue, M. 1994, ApJ, 422, 586
\bibitem[Argon et al.(2004)]{arg04} Argon, A. L. Greenhill, L. J., Moran, J. M., 
Reid, M. J., Menten, K. M., \& Inoue, M. 2004, ApJ 615,  702
\bibitem[Argon et al.(2007)]{arg07} Argon, A. L., 
Greenhill, L. J., Reid, M. J., Moran, J. M., \& Humphreys, E. M. L. 2007, ApJ, 659, 1040
\bibitem[Bahcall et al.(1997)]{bah97} Bahcall, J. N., Kirhakos, S., 
Saxe, D. H., \& Schneider, D. P. 1997, ApJ, 479, 642
\bibitem[Ball et al.(2005)]{bal05} Ball, G. H., Greenhill, L. J., Moran, J. M., Zaw, I., 
\& Henkel, C. 2005, ASPC, 340, 235
\bibitem[Barazza et al.(2008)]{bar08} Barazza, F. D., Jogee, S., \& Marinova, I. 2008, ApJ, 675, 1194
\bibitem[Barth et al.(1999)]{bar99} 
Barth, A. J., Tran, H. D., Brotherton, M. S., Filippenko, A. V., 
Ho, L. C., van Breugel, W., Antonucci, R., 
\& Goodrich, R. W. 1999, AJ, 118, 1609
\bibitem[Barvainis \& Antonucci(2005)]{bar05} Barvainis, R. \& Antonucci, R. 2005, \apjl, 628, 89 
\bibitem[Baudry \& Brouillet(1996)]{bau96}
 Baudry, A., \& Brouillet, N. 1996, A\&A, 316, 188
\bibitem[Becker et al.(1993)]{bec93}
 Becker, R., Henkel, C., Wilson, T. L., \& Wouterloot, J. G. A. 1993, 
A\&A, 268, 483
\bibitem[Bennert et al.(2008)]{ben08} Bennert, N., Canalizo, G., Jungwiert, B., 
Stockton, A., Schweizer, F., Peng, C. Y., Lacy, M. 2008, ApJ, 677, 846
\bibitem[Braatz \& Gugliucci(2008)]{bra08} 
Braatz, J. A., \& Gugliucci, N. E. 2008, ApJ, 678, 96 
\bibitem[Braatz et al.(1994)]{bra94}
Braatz, J. A., Wilson, A. S., \& Henkel, C. 1994, ApJ, 437, L99
\bibitem[Braatz et al.(1996)]{bra96}
 Braatz, J. A., Wilson, A. S., \& Henkel, C. 1996, ApJS, 106, 51 
\bibitem[Braatz et al.(1997)]{bra97} Braatz, J. A., Wilson, A. S., \& Henkel,
  C. 1997, ApJS, 110, 321
\bibitem[Braatz et al.(2003)]{bra03}
 Braatz, J. A., Wilson, A. S., Henkel, C., Gough, R., \& Sinclair, M. 2003, ApJS, 146, 249 
\bibitem[Braatz et al.(2004)]{bra04} Braatz, J. A., Henkel, C, Greenhill, L. J., Moran, J. M., \& Wilson, A. S. 2004, ApJ, 617, L29
\bibitem[Braatz et al.(2007)]{bra07} Braatz et al. 2007, IAUS 242,
Astrophysical Masers and their Environments, eds. J. M. Chapman \& W. A. 
Baan, Cambridge University Press, Cambridge, 399
\bibitem[Brunthaler et al.(2005)]{bru05} Brunthaler, A., 
Reid, M., Falcke, H., Greenhill, L. J., \& Henkel, C. 2005, Sci 307, 1440
\bibitem[Brunthaler et al.(2006)]{bru06} Brunthaler, A.,
Henkel, C., de Blok, W. J. G., Reid, M. J., Greenhill, L. J., \& Falcke, H.
2006, A\&A, 457, 109
\bibitem[Canalizo \& Stockton(2001)]{can01} Canalizo, G., \& Stockton, A. 2001, ApJ, 555, 719
\bibitem[Canalizo et al.(2007)]{can07} Canalizo, G., Bennert, N., Jungwiert, B., 
Stockton, A., Schweizer, F., Lacy, M., \& Peng, C. 2007, ApJ, 669, 801
\bibitem[Castangia et al.(2008)]{cas08} Castangia, P., Tarchi, A., Henkel, C. \& Menten, K. M. 2008, A\&A, 479, 111
\bibitem[Churchwell et al.(1977)]{chu77}
Churchwell, E., Witzel, A., Huchtmeier, W., 
Pauliny-Toth, I., Roland, J., \& Sieber, W. 1977, A\&A, 54, 969 
\bibitem[Claussen et al.(1984)]{cla84}
 Claussen, M., Heiligman, G. M., \& Lo, K. Y. 1984, Nature, 310, 298
\bibitem[Claussen et al.(1998)]{cla98}
 Claussen, M. J., Diamond, P. J., Braatz, J. A., Wilson, A. S., \& Henkel, C. 1998, ApJ, 500, L129
\bibitem[Combes(2006)]{com06} Combes, F. 2006, RMxAC, 26, 131
\bibitem[Condon(1989)]{con89} Condon, J. J. 1989, ApJ, 338, 13
\bibitem[Darling \& Giovanelli(2002)]{dar02} Darling, J. \& Giovanelli, R. 2002, ApJ, 572, 810
\bibitem[Disney et al.(1995)]{dis95} Disney, M. J., et al. 1995, Nature, 376, 150
\bibitem[Dos Santos \& L{\'e}pine(1979)]{dos79}
 Dos Santos, P. M., \& L{\'e}pine, J. R. D. 1979, Nature, 278, 34
\bibitem[Falcke et al.(2000a)]{fal00a} Falcke, H., Henkel, C., Peck, A. B., Hagiwara, Y.,
Prieto, A. M., \& Gallimore, J. F. 2000a, A\&A, 358, L17
\bibitem[Falcke et al.(2000b)]{fal00b} Falcke, H., Wilson, A. S., Henkel, C., Brunthaler, A.,
\& Braatz, J. A. 2000b, ApJ, 530, 13
\bibitem[Floyd et al.(2004)]{flo04} Floyd, D. J. E., Kukula, M. J., Dunlop, J. S., 
McLure, R. J., Miller, L., Percival, W. J., Baum, S. A., \& O'Dea, C. P.  2004, MNRAS,
355, 196
\bibitem[Fullmer \& Lonsdale(1989)]{ful89}
Fullmer. L., \& Lonsdale, C. 1989, 
Cataloged Galaxies and Quasars Observed in the IRAS Survey, Version 2, JPL D-1932
\bibitem[Gallimore et al.(2001)]{gal01} Gallimore, J. F., Henkel, C., Baum, S.A., Glass, I. S.,
Claussen, M. J., Prieto, M. A., \& von Kap-Herr, S. 2001, ApJ, 556, 694 
\bibitem[Gardner \& Whiteoak(1982)]{gar82}
Gardner, F. F., \& Whiteoak, J. B. 1982, MNRAS, 201, 13
\bibitem[Greenhill et al.(2008)]{gre08} Greenhill, L. J., Tilak, A., \& Madejski, G. 2008,
ApJL accepted (arXiv:0809.1108)
\bibitem[Greenhill(2004)]{gre04} Greenhill, L.J. 2004, New Astron. Rev. 48, 1079
\bibitem[Greenhill et al.(1997a)]{gre97a}
 Greenhill, L. J., Herrnstein, J. R., Moran, J. M., Menten, K. M, \& Velusamy, T. 1997a, ApJ, 486, L15
\bibitem[Greenhill et al.(1993)]{gre93} Greenhill, L. J., Moran, J. M., Reid, M. J., Menten, K. M., \& Hirabayashi, H.
1993, ApJ, 406, 482
\bibitem[Greenhill et al.(1995)]{gre95} Greenhill, L. J., Jiang, D. R., 
Moran, J. M., Reid, M. J., Lo, K. Y., \& Claussen, M. J. 1995, ApJ, 440, 619
\bibitem[Greenhill et al.(1997b)]{gre97b} Greenhill, L. J., Ellingsen, S. P.,
Norris, R. P., Gough, R. G., Sinclair, M. W., Moran, J. M., \& Mushotzky, R. 1997b, 
ApJ, 474, L103
\bibitem[Greenhill et al.(2003a)]{gre03a}
Greenhill, L. J., Kondratko, P. T., Lovell, J. E. J., Kuiper, T. B. H., Moran, J. M., Jauncey, D. L., 
\& Baines, G. P. 2003a, ApJ, 582, L11
\bibitem[Greenhill et al.(2002)]{gre02}
Greenhill, L. J., et al. 2002, ApJ, 565, 836
\bibitem[Greenhill et al.(2003b)]{gre03b}
 Greenhill, L. J., et al. 2003b, ApJ, 590, 162
\bibitem[Guyon, Sanders \& Stockton(2006)]{guy06} Guyon, O., Sanders, D. B., \& Stockton, A. 2006, ApJS, 166, 89
\bibitem[Hagiwara et al.(2002)]{hag02}
 Hagiwara, Y., Diamond, P. J., \& Miyoshi, M. 2002, A\&A, 383, 65
\bibitem[Hagiwara et al.(2003a)]{hag03a}
 Hagiwara, Y., Diamond, P. J., \& Miyoshi, M. 2003a, A\&A, 400, 457
\bibitem[Hagiwara et al.(2001a)]{hag01a}
 Hagiwara, Y., Diamond, P. J., Nakai, N., \& Kawabe, R. 2001a, ApJ, 560, 119 
\bibitem[Hagiwara et al.(2001b)]{hag01b}
 Hagiwara, Y., Henkel, C., Menten, K. M., \& Nakai, N. 2001b, ApJ, 560, L37
\bibitem[Hagiwara et al.(1997)]{hag97}
Hagiwara, Y., Kohno, K., Kawabe, R., \& Nakai, N. 1997, PASJ, 49, 171
\bibitem[Hagiwara et al.(2003b)]{hag03b}
 Hagiwara, Y., Diamond, P. J., Miyoshi, M., Rovilos, E., \& Baan, W. A. 2003b, MNRAS, 344, L53
\bibitem[Haschik \& Baan(1985)]{has85}
 Haschick, A. D., \& Baan, W. A. 1985, Nature, 314, 144
\bibitem[Henkel et al.(1986)]{hen86}
Henkel, C., Wouterloot, J. G. A., \& Bally, J. 1986, A\&A, 155, 193
\bibitem[Henkel et al.(2002)]{hen02} Henkel, C., Braatz, J. A., Greenhill,
L. J., \& Wilson, A. S. 2002, A\&A, 394, L23
\bibitem[Henkel et al.(2004)]{hen04} 
Henkel, C., Tarchi, A., Menten, K. M., \& Peck, A. B. 2004, A\&A, 414, 117
\bibitem[Henkel et al.(1998)]{hen98} Henkel, C., Wang, Y. P., Falcke, H., Wilson, A. S.,
\& Braatz, J. A. 1998, A\&A, 335, 463
\bibitem[Henkel et al.(1984)]{hen84} 
 Henkel, C., G\"usten, R., Downes, D., Thum, C., Wilson, T. L. \& Biermann, P. 1984, A\&A, 141, L1
\bibitem[Henkel et al.(2005a)]{hen05a} Henkel, C., Peck, A. B., Tarchi, A. Nagar, N. M., Braatz, J. A., 
Castangia, P., \& Moscadelli, L. 2005a, A\&A, 436, 75
\bibitem[Henkel et al.(2005b)]{hen05b} Henkel, C., Braatz, J. A., Tarchi, A., 
Peck, A. B., Nagar, N. M., Greenhill, L. J., Wang, M. \& Hagiwara, Y. 2005b, Ap\&SS 295, 107 
\bibitem[Herrnstein et al.(2005)]{her05} Herrnstein, J. R., Moran, J. M., 
Greenhill, L. J., \& Trott, A. S. 2005 ApJ, 629, 719
\bibitem[Herrnstein et al.(1999)]{her99} Herrnstein, J. R., et al. 1999, Nature 400, 539
\bibitem[Ho et al.(1987)]{ho87}
 Ho, P. T. P., Martin, R. N., Henkel, C., \& Turner, J. L. 1987, ApJ, 320, 663
\bibitem[Huchtmeier et al.(1978)]{huc78}
 Huchtmeier, W. K., Witzel, A., Kühr, H., Pauliny-Toth, I.I., \& Roland, J. 1978, A\&A, 64, L21
\bibitem[Huchtmeier et al.(1988)]{huc88}
Huchtmeier, W. K., Eckart, A., \& Zensus, A. J. 1988, A\&A, 200, 26
\bibitem[Humphreys et al.(2008)]{hum08} Humphreys, E. M. L., Reid, M. J., 
Greenhill, L. J., Moran, J. M., \& Argon, A. L. 2008, ApJ, 672, 800
\bibitem[Hutchings et al.(1994)]{hut94} Hutchings, J. B., Holtzman, J., Sparks, 
W. B., Morris, S. C., Hanisch, R. J., \& Mo, J. 1994, ApJ, 429, L1
\bibitem[Ishihara et al.(2001)]{ish01}
 Ishihara, Y., Nakai, N., Iyomoto, N., 
Makishima, K., Diamond, P., Hall, P. 2001, PASJ, 53, 2151
\bibitem[Koekemoer et al.(1995)]{koe95} Koekemoer, A. M., Henkel, C., 
Greenhill, L. J., Dey, A., van Breugel, W., Codella, C., \& Antonucci, R.
1995, Nature, 378, 697
\bibitem[Kondratko et al.(2006a)]{kon06a}
Kondratko, P. T., Greenhill, L. J., \& Moran, J. M., 2006, ApJ, 652, 136
\bibitem[Kondratko et al.(2006b)]{kon06b}
Kondratko, P. T., et al. 2006b, ApJ, 638, 100
\bibitem[Labita et al.(2006)]{lab06} Labita, M., Treves, A., Falomo, R., \& Uslenghi, M. 2006,
MNRAS, 373, 551
\bibitem[Lo(2005)]{lo05} Lo, K. Y. 2005, ARA\&A, 43, 625
\bibitem[Malkan et al.(1998)]{mal98} Malkan, M. A., Gorjian, V., \& Raymond, T. 1998, ApJS, 117, 25
\bibitem[Maiolino et al.(1999)]{mai99} Maiolino, R., Risaliti, G., \& Salvati, M. 1999, A\&A 341, L35
\bibitem[McLure et al.(1999)]{mcl99} McLure, R. J., Kukula, M. J., Dunlop, J. S., Baum, S. A.,
O'Dea, C. P., \& Hughes, D. H. 1999, MNRAS, 308, 377
\bibitem[Miyoshi et al.(1995)]{miy95} Miyoshi, M., Moran, J. M., Herrnstein, J. R., 
Greenhill, L., Nakai, N., Diamond, P., \& Inoue, M. 1995, Nature, 373, 127
\bibitem[Morganti et al.(2004)]{mor04} 
Morganti, R., Greenhill, L. J., Peck, A. B., Jones, D. L., \& Henkel, C. 2004, New Astron. Rev. 48, 1195
\bibitem[Nakai et al.(2002)]{nak02}
 Nakai, N., Sato, N., \& Yamauchi, A. 2002, PASJ, 54, L27
\bibitem[Neufeld(2000)]{neu00} Neufeld, D. A. 2000, ApJ, 542, L99
\bibitem[Neufeld et al.(1994)]{neu94} Neufeld, D. A., Maloney, P. R., \& Conger, S. 1994, ApJ, 436, L127
\bibitem[Ott et al.(1994)]{ott94} Ott, M., Witzel, A., Quirrenbach, A.,
Krichbaum, T. P., Standke, K. J., Schalinski, C. J., \& Hummel, C. A. 1994,
A\&A, 284, 331
\bibitem[Peck et al.(2003)]{pec03}
Peck, A. B., Henkel, C., Ulvestad, J. S., Brunthaler, A., Falcke, H., Elitzur, M., Menten, K. M., 
\& Gallimore, J. F. 2003, ApJ, 590, 149
\bibitem[Reid et al.(2008)]{rei08}
Reid, M. J., Braatz, J. A., Condon, J. J., Greenhill, L. J., Henkel, C., \& Lo, K. Y. 2008,	
arXiv:0811.4345v1
\bibitem[Schmidt(1986)]{sch86} Schmidt, M. 1968, ApJ, 151, 393
\bibitem[Schmitt et al.(2002)]{sch02} Schmitt, H. R., Pringle, J. E., Clarke, C. J., Kinney, A. L.
2002, ApJ, 575, 150
\bibitem[Simpson(2005)]{sim05} Simpson, C. 2005, MNRAS, 360, 565
\bibitem[Solomon \& Vanden Bout(2005)]{sol05} 	Solomon, P. M., \& Vanden Bout, P. A. 2005, ARA\&A, 43, 677
\bibitem[Tarchi et al.(2003)]{tar03} Tarchi, A., Henkel, C., Chiaberge, M. \& Menten,
K. M. 2003, A\&A, 407, L33
\bibitem[Tarchi et al.(2002a)]{tar02a}
 Tarchi, A., Henkel, C., Peck, A. B., \& Menten, K. M. 2002a, A\&A, 385, 1049
\bibitem[Tarchi et al.(2002b)]{tar02b}
 Tarchi, A., Henkel, C., Peck, A. B., \& Menten, K. M. 2002b, A\&A, 389, L39
\bibitem[Tarchi et al.(2007)]{tar07} Tarchi, A., Brunthaler, A., Henkel, C., Menten, K. M., 
Braatz, J., \& Wei{\ss}, A. 2007, A\&A, 475, 497
\bibitem[Tran(2001)]{tra01} Tran, H. D. 2001, ApJ, 554, L19
\bibitem[Trotter et al.(1998)]{tro98}
Trotter, A. S., Greenhill, L. J., Moran, J. M., Reid, M. J., Irwin, J. A., \& Lo, K.-Y.
1998, ApJ, 495, 740
\bibitem[Urrutia et al.(2008)]{urr08} Urrutia, T., Lacy M., \& Becker, R. H. 2008, ApJ, 674, 80
\bibitem[Vignali et al.(2003)]{vig03} Vignali, C., Brandt, W. N., \& Schneider, D. P. 2003,
AJ, 125, 433
\bibitem[Veron-Cetty \& Veron(1996)]{ver96} Veron-Cetty, M.-P., \& Veron, P. 1996, ESO Scientific Report, 17, 1
\bibitem[Vestergaard et al.(2008)]{ves08} Vestergaard, M., Fan, X., Tremonti, C. A., Osmer, P. S.,
\& Richards, G. T. 2008, ApJ, 674, L1
\bibitem[Wilkes et al.(1995)]{wilk95} Wilkes, B. J., Schmidt, G. D., Smith, P. S., Mathur, S., 
\& McLeod, K. K.1995, ApJ, 455, 13
\bibitem[Wilson et al.(1995)]{wil95}
 Wilson, A. S., Braatz, J. A., \& Henkel, C. 1995, ApJ, 455, L127 
\bibitem[Wouterloot \& Walmsley(1986)]{wou86}
 Wouterloot, J. G. A., \& Walmsley, C. M. 1986, A\&A, 168, 237
\bibitem[Wright(2006)]{wri06} Wright, E. L. 2006,
PASP, 118, 1711
\bibitem[Zakamska et al.(2004)]{zak04} Zakamska, N. L., Strauss,
M. A., Heckman, T. M., Ivezic, Z., \& Krolik, J. H. 2004, AJ, 128, 1002
\bibitem[Zakamska et al.(2003)]{zak03} Zakamska, N. L., et al. 2003, AJ, 126, 2125
\bibitem[Zakamska et al.(2005)]{zak05} Zakamska, N. L., et al. 2005, AJ, 129, 1212
\bibitem[Zakamska et al.(2006)]{zak06} Zakamska, N. L., et al. 2006, AJ, 132, 1496
\bibitem[Zhang et al.(2006)]{zha06} Zhang, J. S., 
Henkel, C., Kadler, M., Greenhill, L. J., Nagar, N., Wilson, A. S., \& Braatz, J. A.
2006, A\&A, 450, 933
\end{thebibliography}
\end{document}